\documentclass[letter paper, 12pt]{article}
\usepackage{graphicx}
\usepackage{epsf}
\usepackage{amsmath}
\usepackage{here}
\usepackage{setspace}
\usepackage{natbib}
\usepackage{xcolor}
\usepackage{varioref}
 \usepackage{wrapfig}
 \usepackage{threeparttable}
 \usepackage{dcolumn}
  \newcolumntype{d}{D{.}{.}{-1}}
 \usepackage{nomencl}
  \makeglossary
  \usepackage{subfigure}
 \usepackage{subfigmat}
 \usepackage{fancyvrb}
  \fvset{fontsize=\footnotesize,xleftmargin=2em}
 \usepackage{lettrine}

\usepackage{caption}
\usepackage{epstopdf}
\usepackage{lineno}

\begin{document}
\title{ Unsteady Three-dimensional Rotational Flamelets }
 \author{William A. Sirignano \\
  {\normalsize\itshape  Department of Mechanical and Aerospace Engineering }\\
   {\normalsize\itshape  University of California, Irvine, CA 92697}  \\
 }

\newcommand{\hs}{\mbox{\hspace{0.10in}}}
\newcommand{\hsm}{\mbox{\hspace{-0.06in}}}
\newcommand{\be}{\begin{equation}}
\newcommand{\ee}{\end{equation}}
\newcommand{\bea}{\begin{eqnarray}}
\newcommand{\eea}{\end{eqnarray}}

\maketitle

\begin{abstract}
A new unsteady  flamelet model is developed to be used for sub-grid modeling and coupling with the resolved flow description for turbulent combustion. Difficulties with prior unsteady flamelet models are identified.  The model extends the quasi-steady rotational flamelet model which differs from prior models in several critical ways. (i) The effects of shear strain and vorticity  are determined, in addition to normal-strain-rate impacts. (ii) The strain rates and vorticity are determined from the conditions of the environment surrounding the flamelet without  a contrived progress variable. (iii) The flamelet model is physically three-dimensional but reduced to a one-dimensional, unsteady formulation using similarity. (iv)  Variable density is fully addressed in the flamelet model. (v) Non-premixed flames, premixed flames, or multi-branched flame structures are determined rather than prescribed. For both quasi-steady and unsteady cases, vorticity creates a centrifugal force on the flamelet counterflow that modifies the  transport rates and burning rate.   In the unsteady scenario, new unsteady boundary conditions must be formulated to be consistent with the unsteady equations for the rotating counterflow. Eight boundary values on inflowing scalar and velocity properties and vorticity will satisfy four specific relations and therefore cannot all be arbitrarily specified. The temporal variation of vorticity is connected to the variation of applied normal strain rate through the conservation principle for angular momentum.  Limitations on the model concerning fluctuation of the inter-facial plane are identified and conditions under which inter-facial plane fluctuation is negligible are explained. An example of a rotating flamelet counterflow with oscillatory behavior is examined with linearization of the fluctuating variables.

\end{abstract}



\linespread{1.0}

\section{Introduction}

Combustion in high mass-flux chambers  leads to turbulent flow. Thereby, many length and time scales appear in the physics where the smallest scales often cannot be resolved at reasonable cost in computations. In large-eddy simulations (LES), the smaller scales are filtered  allowing affordable computations, while requiring that the essential, rate-controlling, physical and chemical processes occurring on shorter scales  must be modelled.   Proper coupling to the resolved LES flow field is required.

The focus here is on the shear-driven turbulent flames  of practical combustors. Thereby, we address flamelets as originally described by \cite{Williams1975}, namely ``highly sheared small diffusion flamelets". Vorticity has been shown by \cite{Sirignano_2022a, Sirignano_2022b, Sirignano_2022c}, and  \cite{Hellwig2023} to affect burning rates and flammability limits. The goal in this paper is to extend the quasi-steady rotational flamelet model of those publications to cover unsteady behavior.  By our definition here, we exclude laminar flame studies where the flame and the counterflow are fixed between two opposing nozzles. The incoming flows must not be so rigidly constrained to give a fair representation of a laminar flamelet in a field with turbulent combustion. 

Via appropriate scaling for turbulent flows, estimates for the Kolmogorov scales for velocity, length, time, strain rates and vorticity can be obtained. Strained flames (i.e., counterflow flows will not occur at smaller scales because higher strain rates than the Kolmogorov scale are not possible. Flames with thickness much smaller than an eddy size would be wrinkled flames propagating through the eddies rather than flamelets held in capture by an eddy.  Flames with size larger than some eddies would rely on turbulent diffusion to assist propagation. These flames might be large enough to describe well enough at the resolved scale. Our focus is on flamelets captured by the small eddies; thereby, the physics of the flamelet can be approached through laminar modeling. 

The common gases for combustion have kinematic viscosity $\nu$, thermal diffusivity $\alpha$, and mass diffusivity $D$ all very close in magnitudes. Thereby, we expect the viscous layer near the counterflow interface and the diffusion flame thickness to have similar size and to overlap. The diffusion/viscous layer  centers very close to the interface for the diffusion flame. 

The longer-term goal is to develop a flamelet model that will eventually be able to function with Reynolds-averaged Navier-Stokes computation (RANS), unsteady RANS, and LES computations.  We address flames with the diffusive-advective balance at scales near and including the Kolmogorov scale. In addition to scaling the fluid dynamics from the resolved scale to the smaller length scales, rules are needed to scale the gradients of the scalar properties. These gradients will statistically increase as the turbulence wavenumber increases. That portion of the coupling is one-way. The flamelet model would produce a heat release rate based on these inputs. That heat release affects all scales and creates a two-way coupling between the resolved flow and the small eddy scale.  \cite{Sirignano_2022a} made a primitive attempt at coupling using a mixing-length turbulence model for the resolved scale. A major challenge for advancement is to develop improved methods for scaling the gradients of scalar properties such as temperature and mass fractions.

\subsection{Existing Flamelet Theory}

We must understand the laminar mixing and combustion that commonly occur within the smallest turbulent eddies. These laminar flamelet sub-domains experience significant strain of all types: shear, tensile, and compressive. Some important works exist here but typically for either counterflows with only normal strain or simple vortex structures in two-dimensions or axisymmetry and often with a constant-density approximation. The  contributions of \cite{Bilger1976}, \cite{Linan}, \cite{williams2000}, \cite{Peters}, \cite{Pierce},  \cite{Karagozian}, \cite{Marble},  \cite{cetegen1}, and \cite{cetegen2} are noteworthy. The last three works have a two-dimensional vortical field but without the important  stretching of the vortex in the third direction.  \cite{Karagozian} did consider vortex stretching but, with the uniform density approximation, a critical centrifugal acceleration was not recognized. See the recent  overview of the status of flamelet theory presented  by \cite{Sirignano_2022a, Sirignano_2022b, Sirignano_2022c}.  The two-dimensional planar or axisymmetric counterflow configuration has been  foundational for flamelets and can be observed with a  coordinate system based on the principal strain-rate  directions .  Then, the quasi-steady counterflow can be analyzed by ordinary differential equations because the dependence on the transverse coordinate is either constant or linear, depending on the variable.  
Most flamelet approaches use strain rate or scalar dissipation rate at the diffusion-flame reaction zone (point of stoichiometric mixture fraction) as the key input parameter. Or they make an approximation that a priori relates strain rate with scalar dissipation rate across the flamelet. For example, \cite{Peters} ( page 183) obtains a convenient Gaussian shape for scalar dissipation by inconsistently taking constant density in the momentum equation while using Chapman’s approximation for variable density in the scalar equation.  Our approach is that turbulence imposes the strain rate and vorticity on the incoming flow and heat release within the eddy causes density change and the related modification of velocity and strain rate for the flow through the flamelet. So, the vorticity and strain rate of the far approaching fluid should be the inputs to the analysis. Then, through momentum and mass conservation principles, the velocity and strain rates through the flamelet can be determined. The concept clearly parallels the standard for imposing conditions on the scalar properties; namely, values are imposed on temperature and composition for the upstream incoming flow and not in the flame zone.

Our analytical development in

Relatively little literature addresses unsteady flamelets. The existing flamelet literature addresses diffusion flames without little work on unsteady  premixed flamelets or multi-branched flamelets. None of the work addresses effects of stretched vorticity on the flamelet counterflow. \cite{Peters1984} uses mixture fraction $Z$ to replace the spatial coordinate in the counterflow. It is acknowledged that $Z$ depends on both space and time; the substitution removes advection and convection from explicit appearance in the differential equations at the cost of introducing time dependence both explicitly and, through $Z$,  implicitly. 
\cite{Ghoniem1992} addresses the effects of strain rate on ignition, quenching, and extinction.
\cite{Pitschetal1998} uses a flamelet model for unsteady  hydrogen-air combustion. They show radiation is not significant and compare unsteady results with steady results.
\cite{Peters}, \cite{Ihme-See}, and
\cite{Pitschetal1998} transform coordinates from $x, t$ to $Z, \tau = t$ where $Z$ is the mixture fraction. The new time $\tau$ is arbitrarily identified with the motion of the point where $Z$ holds the stoichiometric value.  This is clearly a rough ad hoc approximation which is not accurate in any interesting limit. Specifically, the point where $Z$ maintains stoichiometric value will not move with the local zelocity because its movement is modified by a diffusion term in the governing equation. For example, in the steady-state limit, the point where $Z$ has any specific value (stoichiometric or otherwise) does not move with the local velocity. So, the physics strongly and clearly contradicts the postulate. \cite{Mueller} transforms from the dependence on three space dimensions plus time to a dependence on $Z$ and a contrived progress variable $\lambda$. He obtains relations with as many as seventy terms because the coordinates are no longer orthogonal. Furthermore, the reduction in the number of independent variables implies that more than one point in space and time can have the same values for $Z$ and $\lambda$, resulting in some confusion for interpretations.  Clearly, opportunity and need exists for further study of unsteady flamelet models.

Both \cite{Pitschetal1998} and \cite{Ghoniem1992} use a density-weighted spatial coordinate without needed adjustment of the time derivative, which is caused by the temporal dependence of the density.
\cite{Peters1984} does not acknowledge that the density-weighted coordinate $\eta$ will depend on both $Z$ and time. \cite{Pitsch-Ihme} use a quasi-steady progress-variable formulation to address unsteady turbulent combustion. However, no time derivative appears in the flamelet formulation. Others have treated unsteady combustion in a similar fashion with a quasi-steady flamelet model: \cite{Tuan1}, \cite{Tuan2}, \cite{Tuan3}, \cite{Shadram2021a}, \cite{Shadram2021b}.

There are concerns with the earlier models that neglected vorticity and the resulting centrifugal acceleration on a three-dimensional counterflow with variable density. Irrotational counterflow is not consistent with the fact that the smaller scales in turbulent combustion experience high vorticity magnitudes. 

There is a vast literature on flamelet theory, explaining quasi-steady flamelets. Both premixed and non-premixed flamelets have been studied. We do not attempt a broad review here. The focus is on unsteady diffusion flames. Although the analysis of SEction \ref{flamelet} can be applied to unsteady or steady pemixed, non-premixed, or partially premixed flames, the calculations and conclusions in Sections \ref{dynamics} and \ref{conclusions} will be more narrow.

\subsection{ Orientations of Principal Strain Axes, Vorticity, and Scalar Gradients} \label{orientation}

 On the small length scale, a statistically accurate representation of the relative orientations  of the vorticity vector, scalar gradients, and the directions of the three principal axes for strain rate is sought. 
 \cite{Ashurstetal}, \cite{Nomura1992}, \cite{Nomura1993},  \cite{Boratav1996}, and \cite{Boratav1998} have produced direct numerical simulations (DNS) with useful information for homogeneous sheared turbulence and isotropic turbulence. The vorticity alignment with the intermediate strain direction is most probable, especially in the case with shear. The intermediate strain rate is most likely to be extensive (positive) implying a counterflow configuration. In regions of exothermic reaction and variable density,  vorticity can align with the most tensile strain direction; however, as the strain rates increase, the intermediate direction becomes more favored for alignment with vorticity.  A material interface most probably aligns to be normal to the  compressive normal strain; i.e.,  the scalar gradient and the compressive strain are aligned. Vortex sheets or ribbons are more common than vortex tubes.  Recent forefront DNS  research on turbulent combustion  has focused on other issues than relative alignments of normal strain eigenvectors, vorticity, and scalar gradients. See the review  by  \cite{Driscolletal2020}.

Based on those understandings concerning vector orientations, \cite{Sirignano_2022a, Sirignano_2022b, Sirignano_2022c} extended flamelet theory in a second significant aspect beyond the inclusion of both premixed and non-premixed flame structures; namely, a model was created of a three-dimensional field with both shear and normal strains.  The three-dimensional problem is reduced to a one-dimensional similar form for the counterflow or mixing-layer flow.  The system of ordinary differential equations (ODEs) is presented for the thermo-chemical variables and the velocity components.  The chemical-kinetic model appears as a source term for an ODE. These new findings are very helpful in improving the foundations for flamelet theory and its use in sub-grid modeling for turbulent combustion.

Given  the needed improvements, the aim here is to develop a flamelet model that
 (i) determines directly the effect of shear strain and vorticity on the  flames; (ii) applies directly the local ambient strain rates and vorticity to the flamelet field without the use of a contrived progress variable; (iii) employs a three-dimensional flamelet model; and (iv) considers the effect of variable density, thereby capturing the centrifugal effect; and (v) determines rather than prescribes the existence of non-premixed flames, premixed flames, or multi-branched flame structures; The analysis uses Fickian diffusion and some constant thermophysical properties  to avoid complications in this initial study; however, a clear template will exist for the employment of multi-step kinetics, detailed transport, and variable thermophysical properties in the spirit of the quasi-steady analysis by \cite{Hellwig2023}.

   Section \ref{flamelet} has the development of  a new, unsteady sub-grid flame model that better handles connection with strain and vorticity on the resolved scale, three-dimensional character, and multibranched flame structure. Section \ref{dynamics} presents analysis of some key consequences of the new unsteady model. Concluding comments are made in Section \ref{conclusions}.

\section{Unsteady Rotational  Flamelet Model Formulation}\label{flamelet}

The counterflow configuration is commonly used as the basis for flamelet modeling. It represents what occurs in a combustor where the locally opposing streams caused by normal compressive strain rate create a balance  between diffusion and advection that allows stabilization of a flame. Without the strain rate, diffusion alone would lead to extinction in time. Generally, with two opposing streams, an inter-facial surface is formed. We consider sufficiently low velocities so that the interface is stable. Depending on the momentum of each opposing stream, a velocity for the movement of the interface normal to the inter-facial surface is determined; thus, based on relative velocities to the inter-facial position, the opposing momentum achieve the proper balance.  Our reference-frame origin is fixed to the interface.  Given the densities of the two incoming streams and the (relative) velocity values that lead to the needed balance, the two inflowing mass fluxes are determined. Generally, the ratio of the mass flux of the fuel stream to the mass flux of the incoming oxidizer stream differs from the chemical stoichiometric mass ratio. For the most common fuel / oxidizer combinations in practice, the fuel mass flux exceeds the required amount for stoichiometry. Reaction occurs therefore on the oxidizer side of the interface. Fuel moves more slowly towards the flame because the advection is opposing its diffusion near the flame. The oxidizer moves more swiftly because its diffusion and advection are aligned. While species velocities are identical to the stream velocity at a distance from the flame, substantial adjustment of the species velocity is naturally constructed to achieve stoichiometric proportion for the species mass fluxes near the flame.

One new feature in our treatment involves the presence of vorticity. The vorticity creates a centrifugal effect promoting a minimum value of pressure at the center of the vortex and slowing motion towards the center of the structure. To the contrary, the counterflow promotes a maximum of pressure at the center of that structure and inherently involves motion towards the center in some directions. Thus, the superposition of the two structures raises interesting questions about the competition between two trends and the resulting impacts on residence time for flow through the dual structure. Residence time has important consequences on the flammability of the reacting flow.

We formulate the problem in an unsteady, three-dimensional form. A Galilean transfornmation is used to have a reference frame that moves with the interface of the counterflow with placement of the origin at the stagnation point. (Later, we will discuss whether that location accelerates under unsteady fluctuations.) The following alignments will be assumed. The direction of major compressive principal strain is orthogonal to the vorticity vector direction.  Specifically, the intermediate principal strain direction is aligned with the vorticity while the scalar gradient aligns with the principal compressive strain direction. These assumptions are consistent with the statistical findings of \cite{Nomura1993}.

\subsection{Coordinate Transformation} \label{transformation}

\begin{figure}
  \centering
  \includegraphics[height = 7.0cm, width=0.8\linewidth]{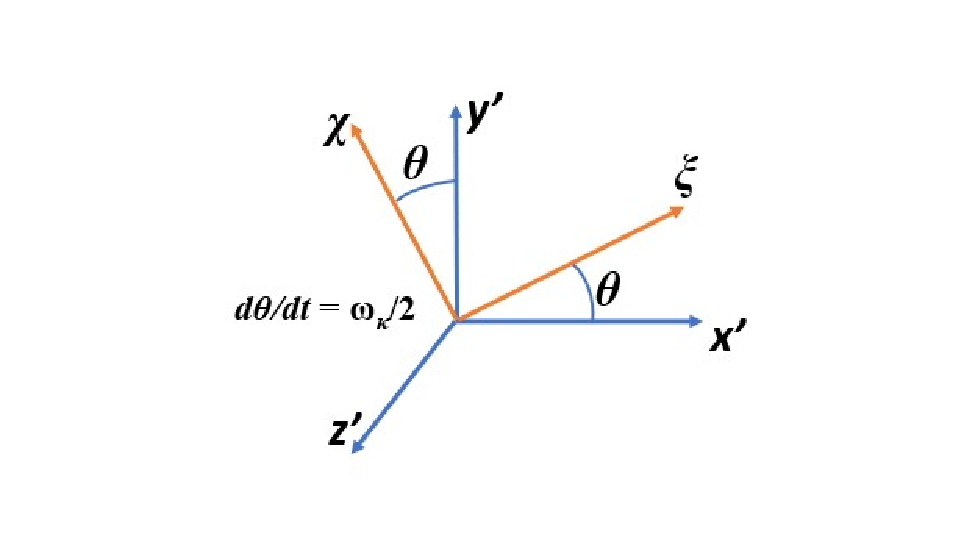}
  \caption{Transformation to $\xi, \chi, z'$ rotating coordinate system from $x', y', z'$ Newtonian system. $\theta$ increases in the counterclockwise direction.   
  }
  \label{Coordinate}
\end{figure}

 A transformation displayed in Figure \ref{Coordinate} is made from the Newtonian frame with rotating material (due to vorticity) to a rotating, non-Newtonian frame.  Let the vorticity direction be the $z'$ direction in an orthogonal framework. Any $x', y'$ plane contains the directions of scalar gradients, major principal axis for compressive strain, and major principal axis for tensile strain.  Note that the $x',y', z'$ directions are not correlated with coordinates on the resolved scale. $\omega$ is the vorticity magnitude on this sub-grid  scale. $x', y', z'$ are transformed to $\xi, \chi, z'$ wherein the material rotation is removed from the $\xi, \chi$ plane by having it rotate at angular velocity $d\theta/dt = \omega/2$ relative to $x', y'$. Here, $\theta$ is the angle between the $x' $ and $\xi$ axes and simultaneously the angle between the $y'$ and $\chi$ axes. We take the sub-grid domain to be sufficiently small to consider a uniform value of $\omega$ across it.

 The following relations apply for the transformation of velocity components $u, v, w$  in the original reference frame to velocity components $u_{\xi}, u_{\chi}, w$ in the rotating reference frame.
\begin{eqnarray}
\xi &=& x' cos \theta + y' sin\theta \;\; ; \;\;
\chi = y' cos \theta - x' sin\theta \nonumber \\
\frac{\partial \xi}{\partial x'} &=&  cos \theta \;\; ; \;\; \frac{\partial \xi}{\partial y'}= sin\theta \;\; ; \;\;
\frac{\partial \chi}{\partial x'} = - sin \theta \;\; ; \;\; \frac{\partial \chi}{\partial y'} = cos \theta \nonumber \\
u_{\xi} &=& u \;cos \theta + v \; sin \theta  + \chi\frac{\omega}{2}   \;\; ; \;\;
u_{\chi} = v\; cos \theta -u \;sin \theta  - \xi\frac{\omega}{2}  \nonumber \\
\frac{\partial u}{\partial x'} &=& \frac{\partial u}{\partial \xi}cos \theta
- \frac{\partial u}{\partial \chi}sin \theta  \;\; ; \;\;
\frac{\partial u}{\partial y'} = \frac{\partial u}{\partial \xi}sin \theta
+ \frac{\partial u}{\partial \chi}cos \theta   \nonumber \\
\frac{\partial v}{\partial x'} &=& \frac{\partial v}{\partial \xi}cos \theta
- \frac{\partial v}{\partial \chi}sin \theta  \;\; ; \;\;
\frac{\partial v}{\partial y'} = \frac{\partial v}{\partial \xi}sin \theta
+ \frac{\partial v}{\partial \chi}cos \theta
\end{eqnarray}
Since
\begin{eqnarray}
\frac{\partial v}{\partial x'}  - \frac{\partial u}{\partial y'} = \omega
\end{eqnarray}
it follows that
\begin{eqnarray}
\frac{\partial u_{\chi}}{\partial \xi}  - \frac{\partial u_{\xi}}{\partial \chi} = 0
\end{eqnarray}
Namely, the flow in the rotating frame of reference does not have the vorticity imposed on it.  However,  two points must be understood. Firstly, the frame is not Newtonian and a reversed (centrifugal) force is imposed. Secondly, the expansions due to combustion and energy release  can produce new vorticity but it will integrate to zero globally.  The inflowing free-stream vorticity in the transformed coordinates is zero.  Thus, in similar fashion to classical counterflow, the vorticity develops as an odd function so that the circulation on a contour surrounding the flow domain remains with zero value. This creation of vorticity is related to gas expansion with density variation; however, that expansion has a symmetry that yields an anti-symmetry in the vorticity.

\subsection{Governing Equations} \label{equations}

The governing equations for unsteady 3D flow in the non-Newtonian frame can be written with $u_i = u_{\xi}, u_{\chi}, w \; ; \; x_i = \xi, \chi, z$. The centrifugal acceleration $a_i = \xi \omega^2/4 , \chi \omega^2 /4, 0$ where $\omega$ may be a function of time ($t$).  The quantities $p, \rho, h, h_m, Y_m, \dot{\omega},  \mu, \lambda, D,$ and $ c_p  $ are pressure, density, specific enthalpy, heat of formation of species $m$, mass fraction of species $m$, chemical reaction rate of species $m$, dynamic viscosity, thermal conductivity, mass diffusivity, and specific heat, respectively. Furthermore, $\tau_{ij} $ is the viscous stress tensor and the Lewis number
$Le \equiv \lambda/(\rho D c_p )$.
\begin{eqnarray}
\frac{\partial \rho}{\partial t} + \frac{\partial (\rho u_j)}{\partial x_j} =0
\label{cont}
\end{eqnarray}
\begin{eqnarray}
\rho \frac{\partial u_i}{\partial t} + \rho u_j\frac{\partial u_i}{\partial x_j} +\frac{\partial p}{\partial x_i} = \frac{\partial \tau_{ij}}{\partial x_j} + \rho a_i
			\label{momentum}
		\end{eqnarray}
		where, following the Stokes hypothesis for a Newtonian fluid,
		\begin{eqnarray}
			\tau_{ij} = \mu \Big[ \frac{\partial u_i}{\partial x_j} + \frac{\partial u_j}{\partial x_i}
			-\frac{2}{3}\delta_{ij} \frac{\partial u_k}{\partial x_k}\Big]
			\label{tau}
		\end{eqnarray}
		\begin{eqnarray}
			\rho \frac{\partial h}{\partial t} + \rho u_j\frac{\partial h}{\partial x_j} - \frac{\partial p}{\partial t} - u_j\frac{\partial p}{\partial x_j}
			=  && \frac{\partial}{\partial x_j} \Big( \frac{\lambda}{c_p} \frac{\partial h}{\partial x_j}  \Big) \nonumber \\
			&&+ \frac{\partial}{\partial x_j} \Big( \rho D (1 - Le)  \Sigma^N_{m=1}h_m \frac{\partial Y_m}{\partial x_j}  \Big) \nonumber \\
			&-&  \rho \Sigma^N_{m=1}h_{f,m} \dot{\omega}_m
			+ \tau_{ij}\frac{\partial u_i}{\partial x_j}
			\label{energy}
		\end{eqnarray}
		\begin{eqnarray}
			\rho \frac{\partial Y_m}{\partial t} + \rho u_j\frac{\partial Y_m}{\partial x_j}  = \frac{\partial }{\partial x_j}\Big(\rho D \frac{\partial Y_m}{\partial x_j}\Big) + \rho \dot{\omega}_m   \;\;;\;\; m=1, 2, ...., N
			\label{species}
		\end{eqnarray}
  Here, we define the non-dimensional  Prandtl, Schmidt, and Lewis numbers:
$ Pr \equiv c_p \mu / \lambda$ ; $Sc \equiv \mu / (\rho D)$  ; and $ Le \equiv Sc/ Pr$.
 Equations (\ref{cont}) through (\ref{species}) together with the equation of state and the relations describing fluid physico-chemical properties give a complete description of behavior in the non-Newtonian reference frame. These equations will be used in the remainder of this section.
		
		Equation (\ref{energy}) is a thermodynamic statement  wherein, following a material element, we are stating the differential relation  $\rho dh - dp = \rho T ds$.  The terms on the right side of (\ref{energy}) are entropy-producing terms.   An alternative form of the energy equation can be developed to govern the total $H$ of the specific enthalpy, specific chemical energy, and kinetic energy per unit mass. That is, $H \equiv h + \Sigma_{m=1}^NY_m h_{f,m} +u_k u_k/2$. Specifically, the vector dot product of $u_i$ with Equation (\ref{momentum}) is used to substitute for $u_j \partial p/ \partial x_j $ in Equation (\ref{energy}) and Equation (\ref{species}) is used to substitute for $\dot{\omega}_m$ there. The Lewis number $Le =1$ is considered. It follows that
		\begin{eqnarray}
			\rho \frac{\partial H}{\partial t} + \rho u_j\frac{\partial H}{\partial x_j} - \frac{\partial p}{\partial t}
			&=&  \frac{\partial }{\partial x_j} \Big( \frac{\lambda}{c_p} \frac{\partial (h+ \Sigma_{m=1}^NY_m h_{f,m}) }{\partial x_j} \Big) \nonumber \\
			&& + \frac{\partial (u_i \tau_{ij})}{\partial x_j}+\rho u_j a_j 
			\label{energy2}
		\end{eqnarray}
		
The energy source term $\rho u_j a_j = \rho (\omega/2)^2(\xi u_{\xi} + \chi u_{\chi})$. If we neglect terms of the  order of the kinetic energy per mass, this effect disappears. Accordingly, we may neglect the viscous dissipation rate $\tau_{ij} \partial u_i/\partial x_j$ . Neglecting kinetic energy, 
$H \approx h + \Sigma_{m=1}^NY_m h_{f,m}$.
The resulting equation becomes
\begin{eqnarray}
			\rho \frac{\partial H}{\partial t} + \rho u_j\frac{\partial H}{\partial x_j} - \frac{\partial p}{\partial t}
			\approx \frac{\partial}{\partial x_j} \Big( \frac{\lambda}{c_p} \frac{\partial H}{\partial x_j}  \Big)
			\label{energy3}
		\end{eqnarray}

		The non-dimensional forms of the above equations remain identical to the above forms if we choose certain reference values for normalization. In the remainder of this article, the non-dimensional forms of the above equations are considered. The superscript $^*$ is used to designate a dimensional property. The variables $u_i^*, t^*, x_i^*, \rho^*, h^*$ (or $H^*$), $p^*,$ and $\dot{\omega}_m^* ,$ and properties $\mu^*, \lambda^*/ c_p^*,$ and $ D^*$ are normalized respectively by the following  quantities specified as constants representative of the magnitudes of the variables:
		$ [(S_1^* +S_2^*) \mu_{\infty}^*/ \rho_{\infty}^*]^{1/2}, (S_1^* + S_2^*)^{-1}, [ \mu_{\infty}^*/ (\rho_{\infty}^*(S_1^* +S_2^*))]^{1/2},
		\rho_{\infty}^*, (S_1^* +S_2^*) \mu_{\infty}^*/ \rho_{\infty}^*, (S_1^* + S_2^*)\mu_{\infty}^*, (S_1^* + S_2^*), \mu_{\infty}^*, \mu_{\infty}^*,$ and $\mu_{\infty}^*/ \rho_{\infty}^*$. The dimensional strain rates $S^*_1$ and $S^*_2$ and  vorticity $\omega^*$ are normalized by $S* \equiv S^*_1 + S^*_2$.  It is understood that, for unsteady flow, the reference values for strain rates and far-stream variables and properties used for normalization  are constants; for example, initial values or average values might be taken for normalization of the  fluctuating variables. Note that the reference length $[ \mu_{\infty}^*/ (\rho_{\infty}^*(S_1^* +S_2^*))]^{1/2}$ is the estimate for the magnitude of the viscous-layer thickness.  In the following flamelet analysis, the vorticity $\omega$ and the velocity derivatives $\partial u_i/ \partial x_j$ are non-dimensional quantities; their dimensional values can be obtained through multiplication by $S* = S^*_1 + S^*_2$.
		
\subsection{Similar Form for the Velocity and Pressure} \label{similar}

		The stagnation point in  the  counterflow  is taken as the origin $\xi = \chi =z=0$. We can make a Galilean transformation to have a stagnation point; however, acceleration of that point must be considered as negligible to obtain the desired similar form. Along the line $\xi=z=0$ normal to the interface, we can expect the first derivatives of $u_{\chi}, \rho, h, T,$ and $Y_m$ with respect to either $\xi$ or $z$ to be zero-valued. In the rotating reference frame, we will examine required conditions for the inter-facial plane between the two incoming streams to remain at $\chi = 0$ and the associated stagnation point to remain at the origin, even with unsteady fluctuations in the ambient conditions.  The velocity components $u_{\xi}$ and $w$ are odd functions of $\xi$ and $z$, respectively, going through zero and changing sign at that line. Consequently, upon neglect of terms of $O(\xi^2)$ and $O(z^2)$, the variables $u_{\chi}, \rho, h, T,$ and $Y_m$ can be considered to be functions only of $t$ and $\chi$. For steady flow, the density-weighted \cite{Illingworth} transformation of $\chi$ can be used to replace $\chi$ with $\eta \equiv \int^{\chi}_0 \rho (\chi') d\chi'$. However, this specific transformation will not be used for the unsteady state since density and $ \eta$ are time dependent. A modified form of a density-weighted variable eliminating time dependence in the coordinate definition will be used in Sub-section \ref{oscillation}.  Note $u_{\chi}$ is independent of $\xi$ and  $z$, $u_{\xi}  \equiv F_1(t, \chi) \xi$ is independent of $z$,  and $w \equiv F_2(t, \chi)z$ is independent of $\xi$ in this case where no shear strain is imposed on the incoming stream(s) after transformation to the rotating coordinate system. At  large positive $\chi$,
		$F_1 \rightarrow F_{1, \infty}(t), F_2 \rightarrow F_{2, \infty}(t).$  Note the relation of the definitions of $F_1$ and $F_2$ here with quantities used in the quasi-steady analyses of \cite{Sirignano_2022a},  \cite{Sirignano_2022b}, and \cite{Hellwig2023}:  $F_1 = S_1 df_1/d\eta, F_2 = S_2 df_2/d \eta$.

      We avoid the common method (e.g., \cite{Pierce}) of using a conserved scalar, typically the mixture fraction $Z$, to replace the spatial coordinate $\chi$ for several reasons.  The premixed flame can have a uniform value of $Z$ over all space of the counterflow. With possible extension in the future to detailed multi-component transport, counter-gradient diffusion might exist locally and thereby it is not guaranteed that $Z$ varies monotonically through the space. Also, the conserved scalar asymptotes to  constant values upstream in both directions for the counterflow; yet, since velocity does not asymptote to a constant, it is  appropriate to have a dependent variable that can capture the variation. In the unsteady case, $Z$ will be a function of time as well as spatial coordinate; accordingly, the time derivative in the reformulation does not equal the original time  derivative.

		In the non-dimensional form given by Equations (\ref{cont}) through (\ref{energy3}), the dimensional strain rates $S_1^*$ and $S_2^*$ are each normalized by the dimensional sum $S_1^* + S_2^*$ at the initial time. Thus, the non-dimensional relation at the initial time is $S_2 = 1 - S_1$; however, for the unsteady case, values of both $S_1(t)$  and $S_2(t)$   must be provided and they generally would not sum to unity.

 From this point forward, for simplicity in expression, we reset $u_{\chi} = v$, although understanding it differs from the $y$-component of velocity in the original Newtonian frame.  The momentum equation  (\ref{momentum}) may be recast as three equations for $v, F_1$, and $F_2$  after substitution with the  relations for velocity components:
		\begin{eqnarray}
		\frac{1}{\xi}\frac{\partial p}{\partial \xi} &=& -\rho \frac{\partial F_1}{\partial t} 
   +\frac{\partial}{\partial \chi}\bigg[\mu \frac{\partial F_1}{\partial \chi}\bigg] - \rho v\frac{\partial F_1}{\partial \chi} -\rho F_1^2 + \rho\bigg(\frac{\omega}{2} \bigg)^2    
       \label{pressureXi}
       \end{eqnarray}
       \begin{eqnarray}
       \frac{\partial p}{\partial \chi} &=&  -\rho \frac{\partial v}{\partial t} 
   + \frac{4}{3}\frac{\partial}{\partial \chi}\bigg[\mu \frac{\partial v}{\partial \chi}\bigg]
   - \rho v\frac{\partial v}{\partial \chi}
   +\frac{\mu}{3}\frac{\partial (F_1 + F_2)}{\partial \chi}  \nonumber \\ 
   && - \frac{2}{3}\frac{\partial \mu}{\partial \chi}(F_1  +F_2)    + \rho \bigg(\frac{\omega}{2} \bigg)^2\chi    \label{pressureChi}
       \end{eqnarray}
          \begin{eqnarray}
	\frac{1}{z}\frac{\partial p}{\partial z} &=&    -\rho \frac{\partial F_2}{\partial t} 
   +\frac{\partial}{\partial \chi}\bigg[\mu \frac{\partial F_2}{\partial \chi}\bigg] - \rho v\frac{\partial F_2}{\partial \chi} -\rho F_2^2 
			\label{pressureZ}
		\end{eqnarray}
  A certain key assumption is implied for obtaining this similar form that reduces the equations  describing the three-dimensional flamelet structure  to depend on only one space dimension. Any other acceleration in the neighborhood of the counterflow interface is assumed to be negligible compared to the centrifugal acceleration. Thus, the stagnation point is considered to remain  fixed  even under the unsteady conditions.

		The following analysis for unsteady counterflow will parallel in some ways the quasi-steady analysis of \cite{Sirignano_2022a, Sirignano_2022b}. In fact, the steady-state limit of this analysis matches exactly the analysis of those papers. It follows from the $\chi$ pressure-gradient in Equation (\ref{pressureChi}) that $\partial p/ \partial \chi$ is a function only of $\chi$ and $t$. Therefore, $\partial^2 p/ \partial \xi \partial\chi = 0$ and $\partial^2 p/ \partial z \partial \chi = 0$. Now, the  right side of the equation for the $\xi$ component of the  pressure-gradient in Equation (\ref{pressureXi}) must be constant  with $\chi$ and a function of $t$ only. The same conclusion is made for the  $z$ component of the  pressure-gradient in Equation (\ref{pressureZ}). The boundary conditions  are  that these $u_{\xi}$ and $w$ velocity components asymptote to values that are independent of $\chi$  as $\chi$   $\rightarrow$ $ \infty$ and $\chi $ $\rightarrow$ $ -\infty$. In the two incoming streams but upstream of the diffusion and reaction zone, the spatial variations of temperature and density will be of the order of the local Mach number squared. Those variations will be neglected here with the assumption that the upstream asymptotes of temperature and density on both streams are constant-valued. However, pressure gradient and velocity can vary upstream. At $\chi =\infty$, $F_1 = F_{1, \infty}(t)$ and  $F_2 = F_{2, \infty}(t)$ which allows the two functions of $t$ to be determined.  Specifically, we obtain
		\begin{eqnarray}
			&& -\rho \frac{\partial F_1}{\partial t} 
   +\frac{\partial}{\partial \chi}\bigg[\mu \frac{\partial F_1}{\partial \chi}\bigg] - \rho v\frac{\partial F_1}{\partial\chi} -\rho F_1^2 + \rho\bigg(\frac{\omega}{2} \bigg)^2 \nonumber \\
   &=& \frac{\partial^2 p}{\partial \xi^2} = g_1(t) = -\rho_{\infty} \frac{\partial F_{1, \infty}}{\partial t} -\rho_{\infty} F_{1, \infty}^2 + \rho_{\infty}\bigg(\frac{\omega}{2} \bigg)^2    \;\;\; ;  \;\;
        \nonumber  \\
			  && -\rho \frac{\partial F_2}{\partial t} 
   +\frac{\partial}{\partial \chi}\bigg[\mu \frac{\partial F_2}{\partial \chi}\bigg] - \rho v\frac{\partial F_2}{\partial \chi} -\rho F_2^2 \nonumber \\  
   &=&  \frac{\partial^2 p}{\partial z^2} = g_2(t) =  -\rho_{\infty} \frac{\partial F_{2, \infty}}{\partial t} -\rho_{\infty} F_{2,\infty}^2 
   \nonumber \\
			\label{PDEs}
		\end{eqnarray}
These two second derivatives of pressure are uniform over all $\xi, \chi$ and $z$ but may vary with time $t$. The second derivative of pressure in the $\chi$ direction can depend on both $\chi$ and $t$. Reorganization of these equations yields partial differential equations for the non-dimensional normal strain rates  $F_1$ and $F_2$:
	\begin{eqnarray}
		\frac{\partial F_1}{\partial t} + v\frac{\partial F_1}{\partial\chi} + F_1^2 &=&
    \frac{1}{\rho}\frac{\partial}{\partial \chi}\bigg[\mu \frac{\partial F_1}{\partial \chi}\bigg]  
   +\frac{\rho_{\infty}}{\rho} \frac{\partial F_{1, \infty}}{\partial t} +\frac{\rho_{\infty}}{\rho}  F_{1, \infty}^2 \nonumber \\  
   && +\bigg[1 -\frac{\rho_{\infty}}{\rho}\bigg] \bigg(\frac{\omega}{2} \bigg)^2   \;\;\; ;  \;\;
        \nonumber  \\
			  \frac{\partial F_2}{\partial t} + v\frac{\partial F_2}{\partial \chi} + F_2^2  &=&
  \frac{1}{\rho} \frac{\partial}{\partial \chi}\bigg[\mu \frac{\partial F_2}{\partial \chi}\bigg] +\frac{\rho_{\infty}}{\rho}\frac{\partial F_{2, \infty}}{\partial t} +\frac{\rho_{\infty}}{\rho}F_{2,\infty}^2 
   \nonumber \\
			\label{PDEs2}
		\end{eqnarray}
The continuity equation in similar form becomes, after defining $f \equiv - \rho v$,
\begin{eqnarray}
\frac{\partial f}{\partial \chi}  =   \frac{\partial \rho}{\partial t} + \rho(F_1 + F_2)
\label{f}
\end{eqnarray}

  Equations (\ref{PDEs2}) are first-order PDEs  in time but second-order in coordinate $\chi$. Two boundary values for each equation will be required at $\chi = \pm \infty$. These will be developed in the next subsection.  Those two PDEs together with 
  Equation (\ref{f}) will produce symmetric solutions in $\chi$-space for $F_1, F_2, \partial v/ \partial\chi,$ and $ p$ and  antisymmetric solutions for $v$ and $\partial p / \partial \chi$ if $\rho$ and $\mu$ are symmetric.  In most situations with the multi-component, reacting flow relevant to combustion, $\rho$ and $\mu$ are not expected to display the symmetry. For the incoming flows at some distance from the interface and reaction zone, the effect of $\mu$ is negligible, although the density can still differ.

  \subsection{Similar Form for the Scalar Variables}

  In the standard counterflow model, the variables $ \rho, Y_m, h$, and $H$ as well as the thermophysical properties $\mu, \lambda, c_p,$ and $D$ will depend only on $\chi$ and $t$. Thereby, the species continuity equations may be written as
\begin{eqnarray}
			\rho \frac{\partial Y_m}{\partial t} + \rho v\frac{\partial Y_m}{\partial \chi}  = \frac{\partial }{\partial \chi}\Big(\rho D \frac{\partial Y_m}{\partial \chi}\Big) + \rho \dot{\omega}_m   \;\;;\;\; m=1, 2, ...., N
			\label{simspecies}
		\end{eqnarray}
and the energy equation becomes
\begin{eqnarray}
			\rho \frac{\partial H}{\partial t} + \rho v\frac{\partial H}{\partial \chi} - \frac{\partial p}{\partial t}
			= \frac{\partial}{\partial \chi} \Big( \frac{\lambda}{c_p} \frac{\partial H}{\partial \chi}  \Big)
			\label{simenergy}
		\end{eqnarray}

\section{Dynamics of Unsteady Rotational Flamelet}\label{dynamics}

  \subsection{Constraints on Unsteady Inflowing Stream Strain Rates} \label{constraints}

  For given functions of $F_{1, \infty}(t)$ and $F_{2, \infty}(t)$, Equations (\ref{PDEs2}) require certain conditions on $F_1$ and $F_2$ at $\chi = - \infty$. Specifically, certain ordinary differential equations in time result.
\begin{eqnarray}
\frac{d F_{1,-\infty}}{dt} +  F_{1,-\infty}^2 &=&  \frac{\rho_{\infty}}{\rho_{-\infty}}  F_{1, \infty}^2 +\bigg[1 -\frac{\rho_{\infty}}{\rho_{-\infty}}\bigg] \bigg(\frac{\omega}{2} \bigg)^2   + \frac{\rho_{\infty}}{\rho_{-\infty}}\frac{d F_{1, \infty}}{d t}         
\label{BCF1}
   \end{eqnarray}
   \begin{eqnarray}
\frac{d F_{2,-\infty}}{dt} +  F_{2,-\infty}^2 &=&  \frac{\rho_{\infty}}{\rho_{-\infty}}  F_{2, \infty}^2   + \frac{\rho_{\infty}}{\rho_{-\infty}}\frac{d F_{2, \infty}}{d t} 
\label{BCF2}
\end{eqnarray}

Through the continuity equation and specification of upstream density on the two inflows of the counterflow, other conditions appear as boundary conditions on the normal strain rate. Specifically,
\begin{eqnarray}
\bigg[\frac{\partial v}{\partial \chi} + F_1 +F_2 \bigg]_{\infty} =  - \frac{1}{\rho_{\infty}} \frac{d \rho_{\infty}}{d t} 
\approx - \frac{1}{\rho_{\infty}} \frac{\partial \rho_{\infty}}{\partial t}
\label{BC3}
\end{eqnarray}
\begin{eqnarray}
\bigg[\frac{\partial v}{\partial \chi} + F_1 +F_2 \bigg]_{-\infty}= - \frac{1}{\rho_{-\infty}} \frac{d \rho_{-\infty}}{d t}  \approx - \frac{1}{\rho_{-\infty}} \frac{\partial \rho)_{-\infty}}{\partial t}
\label{BC4}
\end{eqnarray}

We may consider Equations (\ref{BCF1}, \ref{BCF2}, \ref{BC3}, \ref{BC4}) as providing four constraints on a system of eight boundary values, each possibly a function of time but asymptotically uniform as $\chi \rightarrow \pm \infty$:  $\rho_{\infty}, \rho_{-\infty},  F_{1, \infty},  F_{1, -\infty},   F_{2, \infty},  F_{2, -\infty},$  \\
$\partial v_{\infty}/ \partial \chi,$
and $ \partial v_{-\infty}/ \partial \chi. $  For example, we might specify $\rho_{\infty}(t), \rho_{-\infty}(t), F_{1, \infty}(t),$ and  $  F_{2, \infty}(t) $. 
Then, these four relations can determine $ F_{1, -\infty}(t),   F_{2, -\infty}(t),   \\  \partial v_{\infty}(t)/ \partial \chi, $  and $ \partial v_{-\infty}(t)/ \partial \chi.$  These solutions yield the four needed boundary values to solve PDEs (\ref{PDEs}) to determine $F_{1}(t, \chi)$ and $F_{2}(t, \chi)$.  The density $\rho(t, \chi)$ will be determined through the equation of state and coupled solutions of the scalar PDEs in a similar form which will be derived below based on Equations (\ref{species}, \ref{energy3}). The velocity $v(t, \chi)$ will next be determined using the continuity equation. The $\chi$-momentum equation (\ref{pressureChi}) can be used to determine the functional dependence of pressure on time and $\chi$.
Pressure variation will be discussed in Sub-section \ref{inviscid}.

The analysis of this sub-section and other sub-sections demonstrates the dual methods by which changes at one location of the flow cause changes elsewhere.  The flow model has both the nature of an incompressible flow (with infinite sound speed) and a diffusive medium.
Some changes are instantaneously felt elsewhere. For example, Equations (\ref{BCF1}, \ref{BCF2}, \ref{BC3}, \ref{BC4}) immediately communicate changes between the two upstream flows. On the other hand, Equations (\ref{pressureXi}) through (\ref{PDEs2}) for the dynamics plus Equations (\ref{simspecies}, \ref{simenergy}) for the scalar quantities show that diffusion of momentum, mass, and heat occur causing changes with a time lag.

  The continuity equation (\ref{cont}) may be rewritten as follows:
  \begin{eqnarray}
  \rho v = \rho_0 v_0 - \int^{\chi}_0\rho(F_1 + F_2) d\chi' - \frac{\partial}{\partial t}\int^{\chi}_0 \rho d \chi'   \nonumber  \\
  \label{cont2}
 \end{eqnarray}

Here, under the unsteady condition, we cannot immediately assume $v_0(t) = 0$ continually. It does become zero if the total stress in the $\chi$ direction at $\chi = 0$ maintains zero gradient. Rewriting the $\chi$-momentum equation from Equation (\ref{pressureChi}) to apply strictly at $\chi =0$ at an instant when $v$ is locally linear in $\chi$ and $v(0,t) =0$, we have
\begin{eqnarray}
			\rho \frac{\partial v}{\partial t}   = - \frac{\partial p}{\partial \chi}  
   + \frac{4}{3}\frac{\partial \mu}{\partial \chi}\frac{\partial v}{\partial \chi}
   +\frac{\mu}{3}\frac{\partial (F_1 + F_2)}{\partial \chi}  \nonumber \\ 
			\label{pressure2}
		\end{eqnarray}
  For a symmetric behavior, we expect
\begin{eqnarray}
	 \frac{\partial p}{\partial \chi}  = 0 \;\;;\;\;  \frac{\partial \mu}{\partial \chi}= 0 \;\; ;\;\; \frac{\partial F_1 }{\partial \chi}= \frac{\partial  F_2}{\partial \chi}=0  \nonumber \\			\end{eqnarray}
  at $\chi =0$ and therefore
\begin{eqnarray}
	 \frac{\partial v}{\partial t}(0, t)  = 0  \nonumber \\			
  \end{eqnarray}
For the moment, we will continue the analysis assuming $v(0, t) =0$ for all time of interest. However, some fine points on that subject will be discussed in Sub-section \ref{interface}.

For cases where $\mu, p, \rho$ would vary with $\chi$, symmetry  of the flow might not occur in all details. We can examine the incoming momenta as $\chi \rightarrow \infty$ and $\chi \rightarrow  - \infty$ to determine how a balance occurs. Specifically, we assume the normal strain rates $F_1, F_2, \partial v / \partial \chi$, viscosity $\mu$, and density $\rho$ all asymptote to constants with respect to $\chi$  at plus and minus infinities (although possibly variable with time). Velocity component $v$ and pressure gradient  $\partial p/\partial \chi$ would vary linearly with $\chi$  in the  two ambiences. Note that the vorticity $\omega$  can be time-dependent, but has already been assumed to be uniform.

 Using the divergence of the momentum equation together with Equations (\ref{BCF1}, \ref{BCF2}),  the Laplacian of the pressure can be evaluated with respect to $t$ and  $\chi$ variations as $\chi \rightarrow +\infty$ and  $\chi \rightarrow -\infty$ when both $\rho_{\infty}$ and $\rho_{-\infty}$ are constant with time. More generally, after time differentiation of Equations (\ref{BC3}, \ref{BC4}), it follows that the Laplacian of pressure in the two ambiences will depend on the second time-derivative of the logarithm of density there.  Once $v(\chi, t)$ is determined, we may return to the $\chi$ momentum equation (\ref{pressureChi}) and determine pressure $p(\chi,t)$ by integration.

 First, the case where $\rho_{\infty}(t) = \rho_{-\infty}(t)$ is examined. (Note that density may still vary through the flamelet.) If $F_{1, \infty}(t)$ and $F_{2,\infty}(t)$ are specified functions of time, Equations (\ref{BCF1},\ref{BCF2})  will determine for any value of $\omega$ that  $F_{1,-\infty}(t) = F_{1, \infty}(t)$ and $F_{2,-\infty}(t) = F_{2,\infty}(t)$. Then, Equations (\ref{BC3}, \ref{BC4})  determine that $( \partial v/ \partial \chi)_{-\infty}(t) = ( \partial v/ \partial \chi)_{\infty}(t)  =  -[F_{1, \infty}(t)+F_{2,\infty}(t) + (1/\rho_{\infty})(d \rho_{\infty}/ dt)] $. Thus, the normal strains in the two ambient inflows will always balance each other in opposition. 

 There is a noteworthy issue. In the case where $\rho_{\infty}(t) \neq \rho_{-\infty}(t)$, Equations (\ref{BCF1}, \ref{BCF2}, \ref{BC3}, \ref{BC4}) will not yield unsteady solutions where opposing strain rates are in phase. 
 
 It is useful to examine in our next sub-section behavior in the limiting case of unsteady, incompressible, inviscid rotational counterflow with differing densities in the opposing streams. Realize that the incoming stream conditions are the same in that limit as they are in our rotational flamelet case. Furthermore, analytical relations are more readily developed and provide useful insights.
 
 \subsection{Unsteady, Incompressible, Inviscid Counterflow with Rotation}  \label{inviscid}

Let us  examine the movement of the inter-facial plane and implications for the relation between the two incoming streams for an unsteady, inviscid, incompressible  counterflow where the two flows  possess different densities. Start with the time-dependent stagnation point $\chi_{stag}(t), \xi =0, z =0$ at the origin $\chi=0, \xi =0, z=0$ at the initial time, $t=0$. Using the velocity potential $\Phi(t, \xi, \tilde{\chi}, Z)$, we may write $\nabla^2 \Phi =0$ everywhere since the velocity is divergence-free in this incompressible case. Define $\tilde{\chi} \equiv \chi - \chi_{stag}$. We may expect a snapshot solution to be
\begin{eqnarray}
    \Phi =\bigg( \frac{\partial v}{\partial \chi}\bigg)_{\infty}\frac{\tilde{\chi}^2}{2} + F_{1, \infty}\frac{\xi^2}{2} 
    + F_{2, \infty}\frac{z^2}{2}  \;\;  ;  \;\;   \tilde{\chi} \geq 0   \nonumber \\
     \Phi =\bigg( \frac{\partial v}{\partial \chi}\bigg)_{-\infty}\frac{\tilde{\chi}^2}{2} + F_{1,- \infty}\frac{\xi^2}{2} 
    + F_{2, -\infty}\frac{z^2}{2}  \;\;  ;  \;\;   \tilde{\chi} \leq 0 
    \label{Potential}
\end{eqnarray}
The unsteady Bernoulli equation can be applied along the central  streamline, $\xi = z =0$ with varying $\tilde{\chi}$ and consideration of the centrifugal effect as potential energy.
\begin{eqnarray}
    \frac{p}{\rho_{\infty}} = - \frac{\partial \Phi}{\partial t} -\frac{v^2}{2} 
    +\bigg(\frac{\omega}{2}\bigg)^2\frac{\tilde{\chi}^2}{2}  +f_{\infty}(t)   \;\;  ; \;\; \tilde{\chi} \geq 0  \nonumber \\
      \frac{p}{\rho_{-\infty}} = - \frac{\partial \Phi}{\partial t} -\frac{v^2}{2} 
    +\bigg(\frac{\omega}{2}\bigg)^2\frac{\tilde{\chi}^2}{2}  +f_{-\infty}(t)   \;\;  ; \;\; \tilde{\chi} \leq 0 
    \label{Bernoulli}
\end{eqnarray}
The solution to Laplace's equation gives a snapshot of $\Phi$ at each instant of time. An arbitrary constant may be added and we choose it so that $\Phi$ = 0 at the stagnation point. So, $\partial \Phi/\partial t = 0$ at that point. Then, in order to match pressure at the interface, we must set   $\rho_{\infty} f_{\infty}(t) = \rho_{-\infty} f_{-\infty}(t)  $ . If we use $v = \partial \Phi/ \partial \tilde{\chi}$ and take the second derivative of the two equations (\ref{Bernoulli}), we obtain
\begin{eqnarray}
			-\frac{\partial^2 p}{\partial \tilde{\chi}^2}  &=& 
   \rho_{\infty} \bigg[ \frac{\partial (\partial v/ \partial \tilde{\chi})}{\partial t}  +  \bigg(\frac{\partial v}{\partial \tilde{\chi}}\bigg)^2  -   \bigg(\frac{\omega}{2} \bigg)^2  \bigg] \; ; \; \tilde{\chi} \geq 0\nonumber \\
    -\frac{\partial^2 p}{\partial \tilde{\chi}^2}&=& 
   \rho_{-\infty} \bigg[ \frac{\partial (\partial v/ \partial \tilde{\chi})}{\partial t}  +  \bigg(\frac{\partial v}{\partial \tilde{\chi}}\bigg)^2  -   \bigg(\frac{\omega}{2} \bigg)^2  \bigg]    \; ; \; \tilde{\chi} \leq 0
			\label{pressure5}
		\end{eqnarray}
   Realize that $\partial v/ \partial \chi = \partial v/ \partial \tilde{\chi}$.  It is piecewise constant with $\tilde{\chi}$ and velocity $v$ is piecewise linear with $\tilde{\chi}$ and zero-valued at the origin in the incompressible case. We may later compare  these results for the incompressible case with the results of Equation (\ref{pressure4}) that applies for the asymptotic behaviors in the viscous, reacting, variable-density case.

  The motion of the interface will be defined by $v_0 \equiv d\chi_{stag}/dt$. From Equation (\ref{Potential}),  we obtain, along the line $\xi = 0, z =0$,
  \begin{eqnarray}
      \frac{\partial \Phi}{\partial t}  = -\bigg(\frac{\partial v}{\partial \chi}\bigg)_{\infty}\tilde{\chi} v_0 + 
      \frac{d (\partial v/ \partial \chi)_{ \infty}}{dt}\frac{\tilde{\chi}^2}{2} \; ; \; \chi \geq 0    \; ; \nonumber \\
        \frac{\partial \Phi}{\partial t}  = -\bigg(\frac{\partial v}{\partial \chi}\bigg)_{-\infty}\tilde{\chi} v_0 + 
      \frac{d (\partial v/ \partial \chi)_{ -\infty}}{dt}\frac{\tilde{\chi}^2}{2} \; ; \; \chi \leq 0  
      \label{dPhi}
  \end{eqnarray}
   We may substitute  Equation (\ref{dPhi}) into Equation (\ref{Bernoulli}), then differentiating with respect to $\tilde{\chi}$ and setting the pressure gradient from both sides of the interface to  be continuous. It requires that  $ v_0 =0$ at that instant of time. If the time derivative of the difference between the two inter-facial pressure gradients is set to zero, demanding continued spatial continuity of the pressure gradient in time, it is easily concluded that $dv_0/dt = 0$ and the inter-facial plane remains at $\chi =0$ during all incompressible fluctuations in the surrounding field. Specifically,
   \begin{eqnarray}
     p_{+} &\equiv&  p(t, 0+)  \; ; \; p_{-} \equiv  p(t, 0-)  \; ;\; \nonumber \\
     \bigg(\frac{\partial p}{\partial \chi}\bigg)_{+} & \equiv & \frac{\partial p}{\partial \chi}(t, 0+)      \; ;\; 
     \bigg(\frac{\partial p}{\partial \chi}\bigg)_{-} \equiv  \frac{\partial p}{\partial \chi}(t, 0-)  \nonumber \\
     p_{+} &=& p_{-}  \;  ;  \;   \nonumber \\
     \bigg(\frac{\partial p}{\partial \chi}\bigg)_{+ }   -  \bigg(\frac{\partial p}{\partial \chi}\bigg)_{-}
    & =& \bigg[ \rho_{\infty}\bigg(\frac{\partial v}{\partial \chi}\bigg)_{+}   -   \rho_{-\infty}\bigg(\frac{\partial v}{\partial \chi}\bigg)_{-}   \bigg ]v_0  =  0   \Rightarrow v_0=0   \; ; \nonumber \\
   \frac{d}{dt}\bigg[\bigg(\frac{\partial p}{\partial \chi}\bigg)_{+ }   -  \bigg(\frac{\partial p}{\partial \chi}\bigg)_{-}\bigg]
  & =&  -v_0 \frac{d}{dt}\bigg[ \rho_{\infty}\bigg(\frac{\partial v}{\partial \chi}\bigg)_{+}   -   \rho_{-\infty}\bigg(\frac{\partial v}{\partial \chi}\bigg)_{-}   \bigg ] \nonumber  \\
  & -& \frac{dv_0}{dt} \bigg[ \rho_{\infty}\bigg(\frac{\partial v}{\partial \chi}\bigg)_{+}   -   \rho_{-\infty}\bigg(\frac{\partial v}{\partial \chi}\bigg)_{-}   \bigg ] =0 \Rightarrow \frac{dv_0}{dt} =0 \nonumber \\
  \label{pressure6}
   \end{eqnarray}
   Thus, asymmetry of density does not appear to cause movement of the interface. 
   
Extension to the viscous, incompressible case is possible through use of viscous potential theory (\cite{Joseph}). We consider piecewise constant density $\rho$ and viscosity coefficient $\mu$ and  match 
total normal stress across the the interface: i.e., 
$p_{+}+\mu_{+}(\partial v/\partial \tilde{\chi})_{+} = p_{-} + \mu_{-}(\partial v/\partial \tilde{\chi})_{-}$.
 The same conclusions as for the inviscid case follow: $v_0(t) =0$  and both 
 $\mu\partial v/\partial \tilde{\chi}$  and $\tilde{\chi} = 0$  are piecewise constant. Note there is no problem with a pressure jump if normal stress is balanced at the interface. In both the inviscid and viscous incompressible cases, the discontinuity in the normal strain rate
  $\partial v/\partial \tilde{\chi}$, implies that $F_1$ and $F_2$ will also be discontinuous at $\tilde{\chi} = 0$. The discontinuities in $F_1$ and $F_2$ imply that velocity components $u_{\xi}$ and $w$ will be discontinuous across   $\tilde{\chi} = 0$. This could be acceptable in the inviscid case but viscosity must make the velocities continuous.  Therefore, this mathematical viscous solution with discontinuity in strain rate cannot be accepted on physical grounds. Intuitively, a solution with variable, continuous strain rate and continuous velocities in the viscous flow would not be more likely to cause fluctuation of the interface. Note that  jumps in viscosity and pressure across the interface are still tolerable if normal stress is balanced there.

  It is expected that, with viscous interaction, the low-density stream incoming with the higher values for 
  $\partial v/\partial \tilde{\chi}, F_1,$  and $F_2$ will  be decelerated while the high-density stream incoming with lower values of those normal strain rates will be accelerated such that, at the interface, strain rates on both sides are equal. This adjustment is expected to create a change in pressure gradient near the interface that causes the interface to depart locally from a planar shape and curve away from the stagnation point towards the incoming low-density side.   It is not clear whether this interface-curvature effect will be important in our flamelet case with continuously varying density and viscosity coefficient.
  
   Asymmetries in  diffusion coefficients and other thermo-physical properties might result in motion of the interface.   We shall proceed with reasonable comfort in assuming $v_0 = 0$ and no local interface curvature in the flamelet case. However,  curvature and potential acceleration of the location of the interface are worthy of future detailed attention.


 As shown in Sub-section \ref{constraints}, the two incoming far streams are in phase with each other and, in fact, matching in values for any value of $\omega$ when $\rho_{-\infty} = \rho_{\infty}$. We will examine here the significance of differences in gas densities in the two incoming streams. Specifically, phase differences of the velocity and strain rates for the two streams and differences in the two far-stream pressure fields will be examined.

An important matter for unsteady counterflow is the relation between the two inflowing pressure fields. The question relates to the phase of the two pressure gradients in the two inflows and effectively to the smoothness of the global pressure function over the field.  Based on assumptions already made, we have $\partial^2 p/\partial \xi^2 =\partial^2 p/\partial z^2 \approx f(t) $ or constant in our domain of interest, after neglect of higher-order terms in the Taylor-series expansion of pressure, i.e., $O(\xi^3, \xi^2 z, \xi z^2, z^3) $. Let us consider  $\partial^2 p/\partial \chi^2$ in the domain of interest. Based on experience with simple  configurations, we expect a counterflow to produce a local maximum of pressure with a negative Laplacian for pressure while  a vortex tends to produce a local minimum of pressure with a positive Laplacian for pressure. Thus, we have, in our case, two opposing actions with regard to pressure. By differentiation of Equation (\ref{pressureChi}), we obtain
\begin{eqnarray}
			\frac{\partial^2 p}{\partial \chi^2}(t, \chi)  &=&  -\frac{\partial \rho}{\partial \chi}\bigg[\frac{\partial v}{\partial t} + v\frac{\partial v}{\partial \chi} -\bigg(\frac{\omega}{2} \bigg)^2\chi   \bigg]   \nonumber \\   &&
   -\rho \bigg[ \frac{\partial (\partial v/ \partial \chi)}{\partial t}  +  \bigg(\frac{\partial v}{\partial \chi}\bigg)^2 + v\frac{\partial^2 v}{\partial \chi^2} -   \bigg(\frac{\omega}{2} \bigg)^2  \bigg] \nonumber \\  &&+ \frac{4}{3}\frac{\partial^2}{\partial \chi^2}\bigg[\mu \frac{\partial v}{\partial \chi}\bigg]
   +\frac{\mu}{3}\frac{\partial^2 (F_1 + F_2)}{\partial \chi^2}  -\frac{1}{3}\frac{\partial \mu}{\partial \chi}\frac{\partial (F_1 + F_2)}{\partial \chi} \nonumber \\ 
 && - \frac{2}{3}\frac{\partial^2 \mu}{\partial \chi^2}(F_1  +F_2)   
			\label{pressure3}
		\end{eqnarray}
Consequently, with the gradients of properties $\rho, \mu,$ and gradients of strain rates $ F_1, F_2 , \partial v/\partial \chi$ going to zero value asymptotically as $\chi \rightarrow  \pm 0$, we obtain
\begin{eqnarray}
			-\frac{\partial^2 p}{\partial \chi^2}\bigg|_{\infty}  &=& 
   \rho_{\infty} \bigg[ \frac{\partial (\partial v/ \partial \chi)_{\infty}}{\partial t}  +  \bigg(\frac{\partial v}{\partial \chi}\bigg)^2_{\infty}  -   \bigg(\frac{\omega}{2} \bigg)^2  \bigg] \nonumber \\
    -\frac{\partial^2 p}{\partial \chi^2}\bigg|_{-\infty}&=& 
   \rho_{-\infty} \bigg[ \frac{\partial (\partial v/ \partial \chi)_{-\infty}}{\partial t}  +  \bigg(\frac{\partial v}{\partial \chi}\bigg)^2_{-\infty}  -   \bigg(\frac{\omega}{2} \bigg)^2  \bigg] 
			\label{pressure4}
		\end{eqnarray}
This equation has some important implications. When the various velocity derivatives are comparable in magnitude, the square of the normal compressive strain rate has a factor of four stronger impact than the square of vorticity, explaining how a counterflow survives with vorticity present. Thus, the tendency is that pressure curvature is negative and its magnitude increases (decreases) as normal-strain-rate magnitude increases (decreases) and as vorticity magnitude decreases (increases). The time derivative of normal strain rate is placed so that pressure curvature is lower (higher) than its ultimate value when strain-rate magnitude is increasing (decreasing). The relation between temporal variations of the normal strain rate and vorticity is discussed in Sub-section \ref{omega}.

We  compare  these results of Equation (\ref{pressure4})  for the asymptotic behaviors in the viscous, reacting, variable-density case with the results of Equation (\ref{pressure5}) for the incompressible case.  Clearly, the asymptotic behaviors for the large magnitudes of $\chi$ is identical in the two sets, with the nuance that the density may be a function of time in the variable-density case. We note that a little more sense is made of the time derivatives in the asymptotic behavior through this comparison. 

If $\rho_{\infty}(t) = \rho_{-\infty}(t)$, Equation (\ref{pressure4}) supports the suspicion about the inflowing streams that the two pressure curvatures are identical and the two compressive strain rates are identical. Actually, with identical densities, Equations (\ref{BCF1}, \ref{BCF2}, \ref{BC3}, \ref{BC4}) conclusively yield that the two normal compressive strain rates are identical. Then, Equation (\ref{pressure4}) assures that the two pressure curvatures are identical. These results hold for both steady and unsteady flow, with or without vorticity.

Equation (\ref{pressure4}) has several other  interesting results. For the case where the two ambient densities are equal (whether varying in time or not), we can expect the two incoming velocity gradients $\partial v/\partial \chi$ to fluctuate in phase; also, the magnitudes of $v_{\infty}$ and $v_{-\infty}$ should vary in phase with time (albeit with opposite direction). However, when the two ambient densities differ, there can be phase differences for the two incoming velocities and their strain rates. 

The right sides of Equations (\ref{pressureXi}) and (\ref{pressureZ}) are functions only of $\chi$; thereby, they equal $\partial^2 p/\partial \xi^2$  and $\partial^2 p/\partial z^2$, respectively. Adding those two terms to the right side of Equation (\ref{pressure3}) yields the Laplacian of the pressure.

\subsection{Implications for the Steady Rotational Flamelet}

The prior unsteady analysis has also yielded some novel results for the steady, rotational flamelet. Based on Equations (\ref{BCF1}, \ref{BCF2}, \ref{BC3}, \ref{BC4}), the following boundary conditions apply for the steady incoming streams:
\begin{eqnarray}
    F_{1,-\infty}^2 &=&  \frac{\rho_{\infty}}{\rho_{-\infty}}  F_{1, \infty}^2 +\bigg[1 -\frac{\rho_{\infty}}{\rho_{-\infty}}\bigg] \bigg(\frac{\omega}{2} \bigg)^2 \;\; ;  \;\; \nonumber \\       
  F_{2,-\infty}^2 &=&  \frac{\rho_{\infty}}{\rho_{-\infty}}  F_{2, \infty}^2  \;\;   ;   \;\; \nonumber  \\
  \bigg[\frac{\partial v}{\partial \chi} + F_1 +F_2 \bigg]_{\infty} &=& 0  \;\;  ;  \;\; \nonumber \\
\bigg[\frac{\partial v}{\partial \chi} + F_1 +F_2 \bigg]_{-\infty} &=&0   \;\;  ;  \;\;  
   \label{steady}
\end{eqnarray}

In the steady-state without vorticity, Equations  (\ref{steady}) yield
\begin{eqnarray}
  \sqrt{  \frac{\rho_{\infty}}{\rho_{-\infty}} } = \frac{F_{1, -\infty}}{F_{1, \infty}}= \frac{F_{2, -\infty}}{F_{12, \infty}} 
  = \frac{\big(\frac{\partial v}{\partial \chi}\big)_{-\infty}}{\big(\frac{\partial v}{\partial \chi}\big)_{\infty}}
\end{eqnarray}
Here, the two ambient values of pressure curvature become equal. However, with vorticity, they only are equal when the two ambient densities are equal.  More generally, with both vorticity and differing ambient densities and after neglect of higher-order terms in $\xi$ and $z$, the steady-state result for the Laplacian of pressure  is that
\begin{eqnarray}
\nabla^2p_{-\infty} - \nabla^2p_{\infty} &=&  2\rho_{\infty} F_{1, \infty} F_{2, \infty}  - 2\rho_{-\infty} F_{1, -\infty} F_{2, -\infty} \nonumber \\
&=& 2\rho_{\infty} F_{1, \infty} F_{2, \infty} \bigg[ 1-    \big[ 1 +\frac{\rho_{-\infty} - \rho_{\infty}}{ \rho_{\infty}}  
\big(\frac{\omega}{2 F_{1, \infty}}\big)^2 \big]^{1/2}    \bigg]  \nonumber \\
&=& 2\rho_{\infty} F_{2, \infty} \frac{\rho_{-\infty} - \rho_{\infty}}{2 \rho_{\infty}F_{1, \infty}}  
\big(\frac{\omega}{2}\big)^2 + O(  \omega^4/ F_{1, \infty}^2 )
\end{eqnarray}
Thus, asymmetry in the steady-state pressure curvature of the ambient fields is caused only by the combination of vorticity and density difference.  The field near the counterflow interface can experience asymmetry due to variation in the viscosity coefficient or asymmetry in the heat release rate.

In setting boundary conditions, the theory applies them at infinite distances in the approaching flows for both the steady case and the unsteady case. For the non-premixed flame, stoichiometry determines on which side - fuel or oxidizer - the diffusion flame is centered. For most practical fuel-oxidizer situations, a small favoring towards the oxidizer side of the interface occurs. Still, the diffusion layer of the flame commonly crosses the interfacial plane.  The flame thickness is estimated as $\delta_f =\sqrt{D/S_f}$ where $S_f$ is the strain rate in the flame region. $S$ is the strain rate applied at a distance on the incoming flow. Typically, $S_f > S$.  We expect the kinematic viscosity $\nu$ and the mass diffusivity $D$ to have the same order of magnitude. Thereby, the thickness of flame diffusive layer $\delta_F \approx \sqrt{\nu/S_f} = \delta$, the viscous layer thickness at the counterflow interface.

\subsection{Movement of the Inter-facial Plane}\label{interface}

We should address possible movement in spatial position of the interface and the  stagnation point in the counterflow  for the unsteady case, especially when the two ambient streams have different densities. This issue exists even if the two ambient densities are constant  and only the velocity field in the ambience is fluctuating. 

For convenience, we define
\begin{eqnarray}
    \mu_{\chi, 0} \equiv  \frac{\partial \mu}{\partial \chi}(t, 0) \;\; ; \;\;  
      \rho_{\chi, 0} \equiv  \frac{\partial \rho}{\partial \chi}(t, 0)  \;\;  ; \;\;  
      p_{\chi,0} \equiv  \frac{\partial p}{\partial \chi}(t, 0)   \;\;  ;  \nonumber \\
       F_{1, \chi, 0} \equiv  \frac{\partial F_1}{\partial \chi}(t, 0) \;\; ; \;\;  
      F_{2, \chi, 0} \equiv  \frac{\partial F_2}{\partial \chi}(t, 0) 
\end{eqnarray}
For the inviscid case or the case with  uniform $\mu$ and uniform $\rho$  values,  Equations (\ref{pressureXi}, \ref{pressureChi}, \ref{pressureZ}, \ref{pressure3}) collectively allow a solution where $\mu_{\chi, 0} =\rho_{\chi, 0}= p_{\chi, 0}
=F_{1, \chi, 0} = F_{2\chi, 0} =   v(t, 0) =0$.  Even where density and viscosity vary through the domain, the smoothness condition described by Equation (\ref{pressure3}) presents a symmetric pressure field and an anti-symmetric pressure-gradient field around  the origin at $\chi = 0$. Thus, $p_{\chi,0} =0$ does not rely on other imposed conditions. If $\mu_{\chi,0}$ and $\rho_{\chi,0}$ hold as local conditions  with gradients of those properties existing elsewhere in the field, the identified equations still support a solution with  $F_{1, \chi, 0} = F_{2,\chi, 0} =   v(t, 0) =0$.

There is a basis for the following argument: if the magnitudes of $\mu_{\chi,0}$ and $\rho_{\chi,0}$ are small compared to values of $F_1, F_2,$  and $\partial v/ \partial \chi$ in the neighborhood of the origin, the assumption $v(t, 0) =0$ is acceptable.  Otherwise, a more complicated configuration exists.  We proceed here with the approximation that $v(t, 0) =0$ is acceptable.

\subsection{Flamelet in Oscillating Environment} \label{oscillation}

An example is needed to demonstrate consequences of temporal variation on the rotating counterflow  flamelet. Although the theory has been developed to address a range of flame structures  (diffusion, premixed, and partially premixed), only one type will be examined here in a computational example. Spefically, a  case for a diffusion flamelet is chosen here with  dynamic equilibrium caused by oscillating pressure and strain rate in the ambient field of the flamelet. These ambient oscillations produce oscillations of both scalar variables and velocity components throughout the flamelet flow field. We assume that oscillations are isentropic and strain rate is 90 degrees out of phase with pressure for the ambient incoming flows. The composition of the incoming streams will not fluctuate. These choices are consistent with asymptotic disappearance of dissipation and constraints via the unsteady continuity equation at large distances from the flame.  Strong dissipation will  appear close to the interface of the flow. The analytical approach is based on control of the flamelet behavior by the conditions of the two incoming streams.

The flow variables will be considered as the sum of a mean, time-independent (i.e., steady-state) term, denoted by superscript bar, and a fluctuating perturbation term, denoted by a superscript prime. Thus, we have  $h(t, \chi) = \bar{h}(\chi) + h'(t, \chi)$, $v(t, \chi) = \bar{v}(\chi) + v'(t, \chi)$, and similar separations for the several other variables. The theory has been developed in earlier sections to capture the full nonlinearities of the flow. The solutions for $\bar{h}(\chi)$ and $\bar{v}(\chi)$ will capture the steady-state behavior, in similar fashion to prior analyses  of \cite{Sirignano_2022a, Sirignano_2022b, Sirignano_2022c}.  The unsteady perturbations will be considered sufficiently small compared to steady-state solutions to yield linearized partial differential equation for the primed variables. Transient behavior will not be examined but rather just the periodic solution associated with the dynamic equilibrium is examined. Thereby, the variable forms become $h' = \hat{h}(\chi)e^{i\sigma t} =  (h_R(\chi) + ih_I (\chi)) e^{i\sigma t}$ and $v' = \hat{v}(\chi)e^{i\sigma t} =  (v_R(\chi) + iv_I (\chi)) e^{i\sigma t}$, where $i$ is the imaginary unit and $\sigma$ is the common frequency for all variables throughout all parts of the flow.  Ultimately, our interest will focus on real part only, yielding perturbations of the forms $h_R cos (\sigma t) - h_I sin (\sigma t)$ and  $v_R cos (\sigma t) - v_I sin (\sigma t)$.  The space-dependent coefficients of the type $h_R, h_I, v_R,$ and $v_I$ will be governed by ordinary differential equations (ODEs). The steady-state ODEs must first be solved to provide inputs for the perturbation ODEs.

In our analysis, we follow many choices of \cite{Sirignano_2022a, Sirignano_2022b, Sirignano_2022c}. Specifically, we assume a one-step chemical reaction for propane and oxygen; equal mass and thermal diffusivities; constant Prandtl number; a perfect-gas equation of state; a constant specific heat; and a constant product of viscosity and density. (This last assumption is somewhat bolder for the unsteady state than for the steady state, but we do accept it.)  The $\chi$ independent variable is transformed to a density-weighted variable, but the steady-state density is used in the transformation avoiding the awkardness, that already appears in the literature, of a hidden time dependence on a spatial coordinate. Thus, $\eta(\chi) \equiv \int_0^{\chi} \bar{\rho} d\chi'$.

For convenience, we define certain differential operators
\begin{eqnarray}
J(\bar{\psi})&\equiv& \frac{d^2 \bar{\psi}}{d\eta^2} +  \bar{f}\frac{d\bar{\psi}}{d \eta} - \bar{\psi}^2
\nonumber \\
L(\bar{\psi})&\equiv& \frac{d^2 \bar{\psi}}{d\eta^2} + Pr \bar{f}\frac{d\bar{\psi}}{d \eta}
\nonumber \\
M(\hat{\psi}) &\equiv& \frac{d^2 \hat{\psi}}{d\eta^2}
+  \bar{f}\frac{d\hat{\psi}}{d \eta}-i  \sigma \hat{\psi} -2 \bar{\psi}\hat{\psi}
 \nonumber \\
N(\hat{\psi}) &\equiv& \frac{d^2 \hat{\psi}}{d\eta^2}
+ Pr \bar{f}\frac{d\hat{\psi}}{d \eta}-i Pr \sigma \hat{\psi} 
\label{operators}
\end{eqnarray}
 Here, $f \equiv - \rho v = \bar{f} + \hat{f}exp(i \sigma t)$.

Consider $\dot{\omega} = \bar{\dot{\omega}}_m  +  \hat{\dot{\omega}}_m exp(i \sigma t)$.
We take Equations (\ref{PDEs2}, \ref{simspecies}, \ref{simenergy} ) to obtain ODEs for both the steady-state variables and the perturbation variables. Equation (\ref{f}) is used in an integrated form.

The governing steady-state equations become
\begin{eqnarray}
 J(\bar{F}_1) +\bigg(\frac{\omega}{2}\bigg)^2   & =& \frac{1}{\bar{\rho}}\bigg[ \bigg(\frac{\omega}{2}\bigg)^2 - \bar{F}^2_{1, \infty} \bigg] 
 \nonumber \\
  J(\bar{F}_2)    & =& - \frac{1}{\bar{\rho}}\bar{F}^2_{2, \infty} 
\nonumber \\
 L(\bar{Y}_m) + Pr\bar{\dot{\omega}}_m &=& 0 \; ; \; m=1, 2, .., N \nonumber \\
  L(\bar{H}) &=& 0   \nonumber  \\
  \bar{f}  &=& \int_0^{\eta}  (\bar{F}_1 + \bar{F}_2) d \eta'
\end{eqnarray}
Boundary conditions are specified for $\bar{F}_{1, \infty} = \bar{F}_{2, \infty} = 0.5,  \bar{Y}_{O, - \infty} = \bar{Y}_{F, \infty} = 1.0,  
 \bar{Y}_{F, - \infty} = \bar{Y}_{O, \infty} = 0,\bar{H}_{-\infty} = 15.75$, and $\bar{H}_{\infty} = 1.0.$  Our example follows  \cite{Sirignano_2022a, Sirignano_2022b, Sirignano_2022c} using oxygen (O) and propane (F).

Then, we may calculate the following:
\begin{eqnarray}
    \bar{F}_{1, -\infty} &=& \sqrt{\bigg(\frac{\omega}{2}\bigg)^2 +\frac{1}{\bar{\rho}_{-\infty} }\bigg[ \bar{F}_{1, \infty}^2 - \bigg(\frac{\omega}{2}\bigg)^2 \bigg] }  \nonumber \\
     \bar{F}_{2, -\infty} &=& \sqrt{\frac{1}{\bar{\rho}_{-\infty} } } \bar{F}_{2, \infty}
     \label{steadyODEs}
\end{eqnarray}

The average nondimensional \cite{Westbrook_Dryer:1984} reaction rate is given by  
$\bar{\dot{\omega}}_F = K Da_{ref}\bar{\rho}^{0.75} \bar{Y}_{F}^{0.1} \bar{Y}_O^{1.65} exp(-\varepsilon/ \bar{h})$ where $\varepsilon = E/RT_{\infty}$.  $E, R, Da,$ and $ K$ are respectively the activation energy, universal gas constant, Damk{\"o}hler number (using a dimensional ambient strain rate $S^* =10,000/second$ and a pressure of ten atmospheres), and a constant allowing adjustment of the applied strain rate and/or pressure.

After separation of the time-dependent portion of the  perturbations, we have
\begin{eqnarray}
 M(\hat{F}_1) &=& \frac{\hat{\rho}}{\bar{\rho}}\frac{d^2 \bar{F}_1}{d\eta^2} +\bigg[\frac{d (\hat{\rho}/\bar{\rho})}{d \eta} - \hat{f}   \bigg]\frac{d \bar{F}_1}{d \eta}  + \frac{\hat{\rho}\bar{F}_1^2- \hat{\rho}_{\infty}\bar{F}_{1,\infty}^2}{\bar{\rho}} - (i\sigma + 2 \bar{F}_{1,\infty})\frac{\bar{\rho}_{\infty}}{\bar{\rho}}\hat{F}_{1,\infty}   \nonumber  \\
 &&+\frac{\hat{\rho}_{\infty} - \hat{\rho}}{\bar{\rho}}\bigg(\frac{\omega}{2}\bigg)^2
 \nonumber  \\
  M(\hat{F}_2) &=& \frac{\hat{\rho}}{\bar{\rho}}\frac{d^2 \bar{F}_2}{d\eta^2} +\bigg[\frac{d (\hat{\rho}/\bar{\rho})}{d \eta} - \hat{f}   \bigg]\frac{d \bar{F}_2}{d \eta}  + \frac{\hat{\rho}\bar{F}_2^2- \hat{\rho}_{\infty}\bar{F}_{2,\infty}^2}{\bar{\rho}} - (i\sigma + 2 \bar{F}_{2,\infty})\frac{\bar{\rho}_{\infty}}{\bar{\rho}}\hat{F}_{2,\infty} 
  \nonumber \\
   N(\hat{Y}_m) &=& - Pr\hat{\dot{\omega}}_m + 2 \frac{\hat{\rho}}{\bar{\rho}}\frac{d^2 \bar{Y}_m}{d\eta^2} 
   +\bigg[\frac{d (\hat{\rho}/\bar{\rho})}{d \eta} 
  + Pr(\bar{f}\frac{\hat{\rho}}{\bar{\rho}}-\hat{f} )\bigg]\frac{d \bar{Y}_m}{d \eta}    \; ; \; m= 1,2, .. , N
  \nonumber \\
    N(\hat{H}) &=& -iPr\frac{\sigma\hat{p}}{\bar{\rho}} +  \frac{\hat{\rho}}{\bar{\rho}}\frac{d^2\bar{H}}{d\eta^2}   +\bigg[\frac{d (\hat{\rho}/\bar{\rho})}{d \eta} -Pr\hat{f}  \bigg]\frac{d \bar{H}}{d \eta}   \nonumber \\
     \hat{f} &=& \int_0^{\eta}  (\hat{F}_1 + \hat{F}_2) d \eta'   +i\sigma\int_0^{\eta} \frac{\hat{\rho}}{\bar{\rho}}d \eta'
    \label{perturbationODEs}
\end{eqnarray}

Boundary conditions are presented as $\hat{F}_{1, \infty} = \hat{F}_{2, \infty} = - 0.05i,  \Hat{}{Y}_{O, - \infty} = \hat{Y}_{O, \infty} =  
 \hat{Y}_{F, - \infty} = \bar{Y}_{F, \infty} = 0$. A pressure amplitude of one atmosphere (ten percent of the mean value) is chosen throughout the field.  $\bar{H}_{-\infty},\bar{H}_{\infty}, \bar{\rho}_{-\infty}$ and $\bar{\rho}_{\infty}$ have isentropic variation with the pressure using the specific-heat ratio $\gamma = 1.4$. Linearized forms of Equations (\ref{BCF1}, \ref{BCF2}) yield the values for $\hat{F}_{1, -\infty}$ and $\hat{F}_{2, -\infty}$. The perturbed reaction rate is \\
 $\hat{\dot{\omega}}_F = \bar{\dot{\omega}}_F [ (\varepsilon/ \bar{h})\hat{p}/\bar{p}  + (0.75 - \varepsilon/ \bar{h})\hat{\rho}/\bar{\rho}  + 0.1 \hat{Y}_F/ \bar{Y}_F + 1.65 \hat{Y}_O/\bar{Y}_O]$.

 Values of the strain rate $\partial v/ \partial \chi$ are not needed for the calculations but may be determined at the boundaries in post-processing with Equations (\ref{BC3}, \ref{BC4}). It may be determined throughout the field using differentiation of $v$ which can be determined from solutions for $f$ and $\rho$.

We have a system involving coupled second-order ODEs, each  with two boundary values. The steady-state equations are nonlinear and provide coefficients for the linear perturbation equations. The perturbation equations actually each form two coupled equations after separation into real and imaginary parts. The numerical method uses iteration with a pseudo-time to produce a converged solution. The  major parameters are $\omega, \sigma, S_1 = 1 - S_2,$ and $ Da = K Da_{ref}.$ The analysis presented to this point can be applied for premixed flames, partially premixed flames, or non-premixed (i.e., diffusion) flames.  We limit the calculations to diffusion flames by choosing pure propane inflow from $\chi = +\infty$ and pure oxygen inflow from $\chi = - \infty$. We choose the oxygen inflow temperature to be $T_{-\infty} = 600K$ while the propane inflow has $T_{\infty} = 300K$.

\begin{figure}[thbp]
  \centering
 \subfigure[Steady-state enthalpy, $\bar{h}$]{
  \includegraphics[height = 4.4cm ]{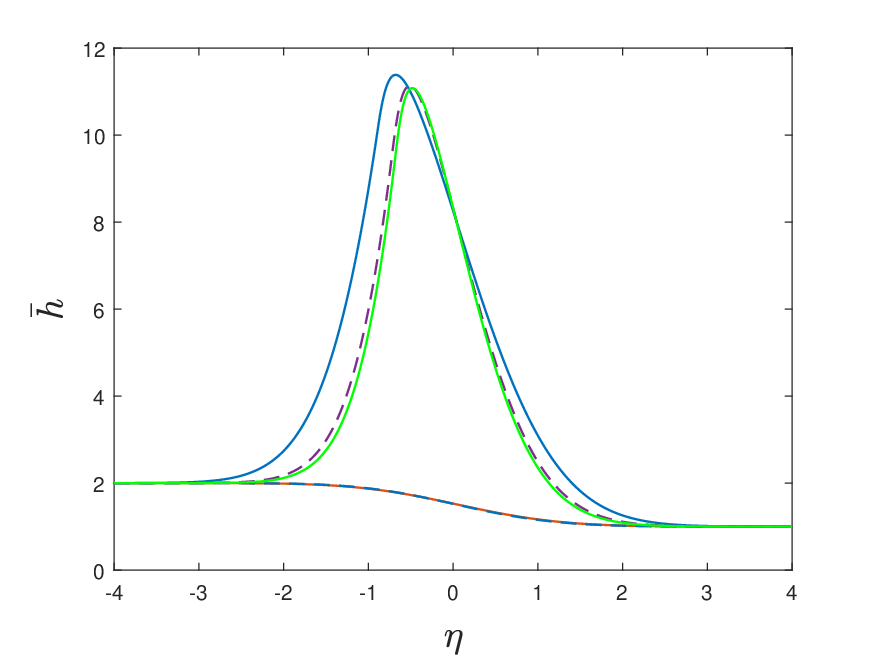}}
  \subfigure[Integrated steady-state mass burning rate, $\int_{-\infty}^{\eta} \bar{ \dot{\omega}}_F d \eta'$]{
  \includegraphics[height = 4.4cm   ]{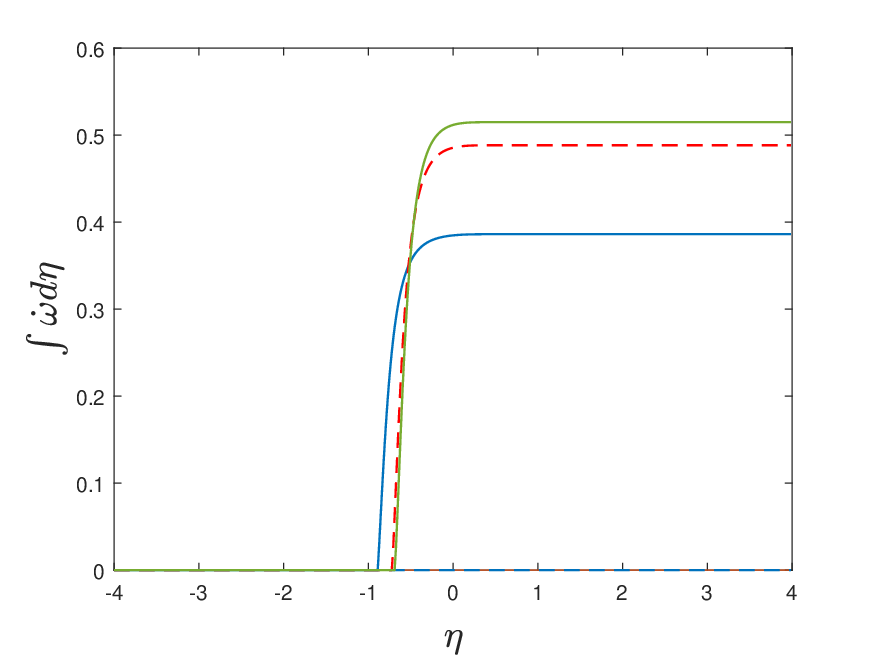}}     \\
  \subfigure[Steady-state mass flux/ area, $\bar{f}(\eta) = - \bar{\rho}\bar{v}$]{
  \includegraphics[height = 4.4cm]{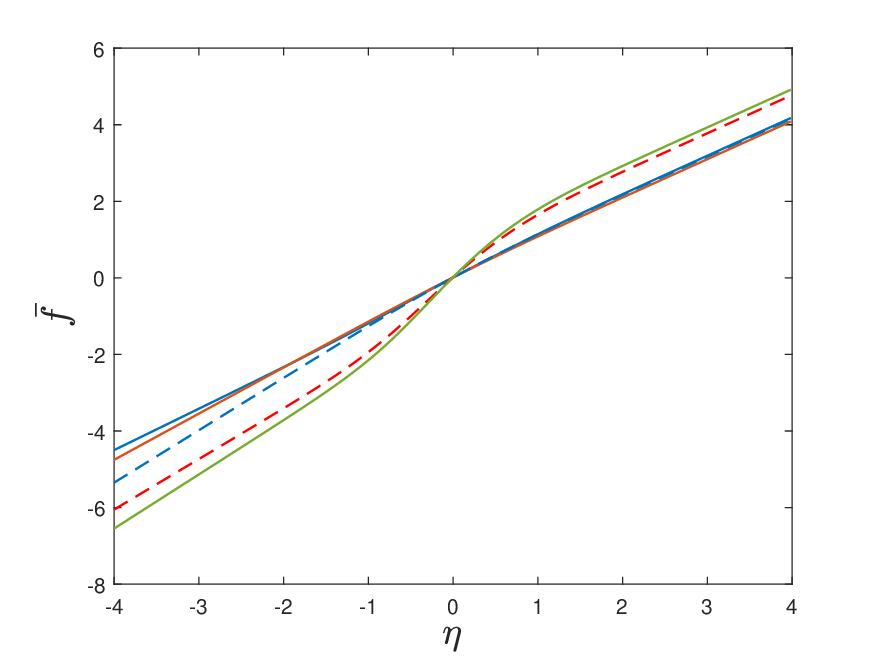}}
  \subfigure[Steady-state velocity, $v(\eta)$]{
  \includegraphics[height = 4.4cm ]{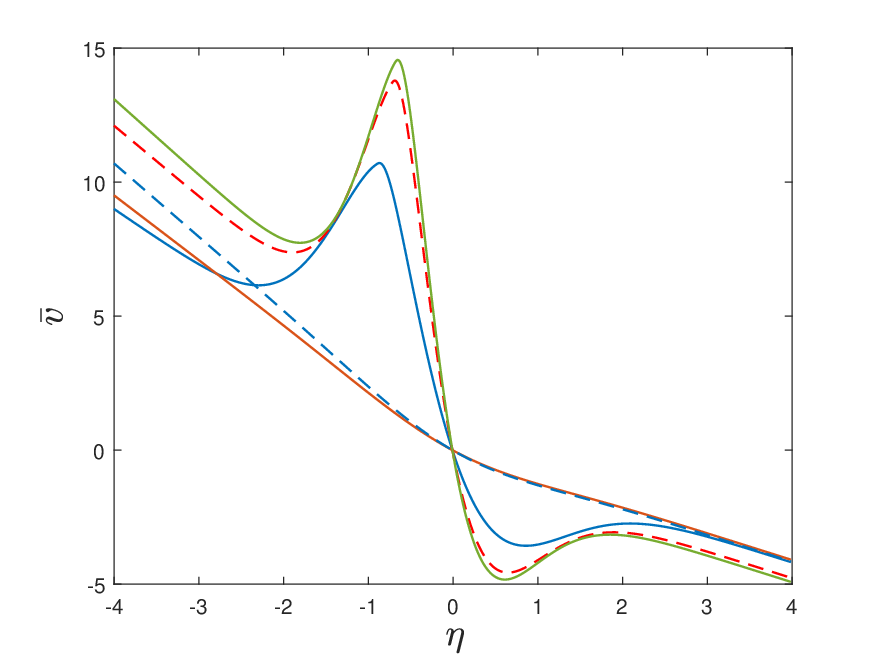}}  \\
    \subfigure[Steady-state normal strain rate $\bar{u}_{\xi}/(S_1*\xi)$]{
  \includegraphics[height = 4.4cm]{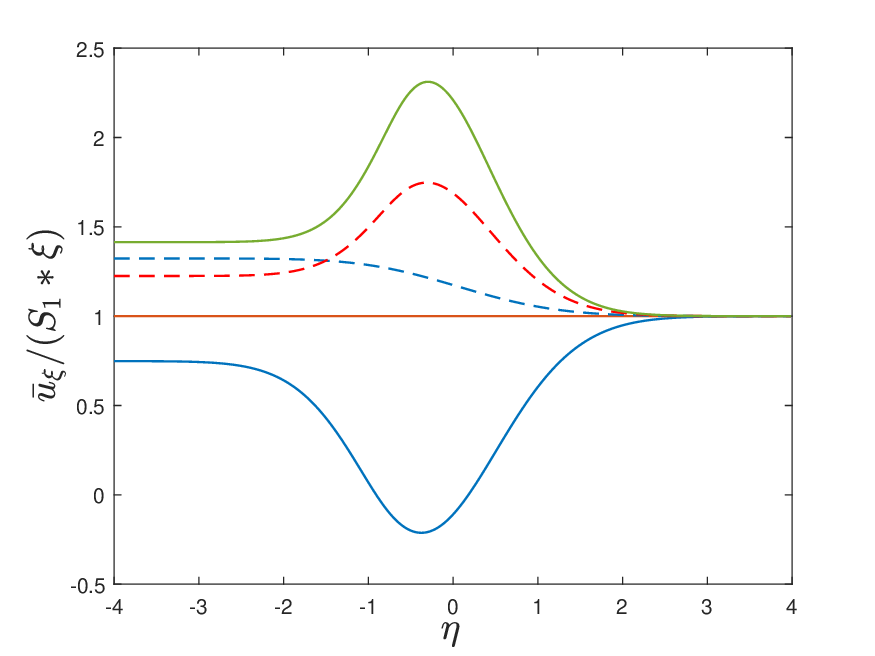}}
   \subfigure[Steady-state normal strain rate $\bar{u}_z/(S_2*z)$]{
  \includegraphics[height = 4.4cm ]{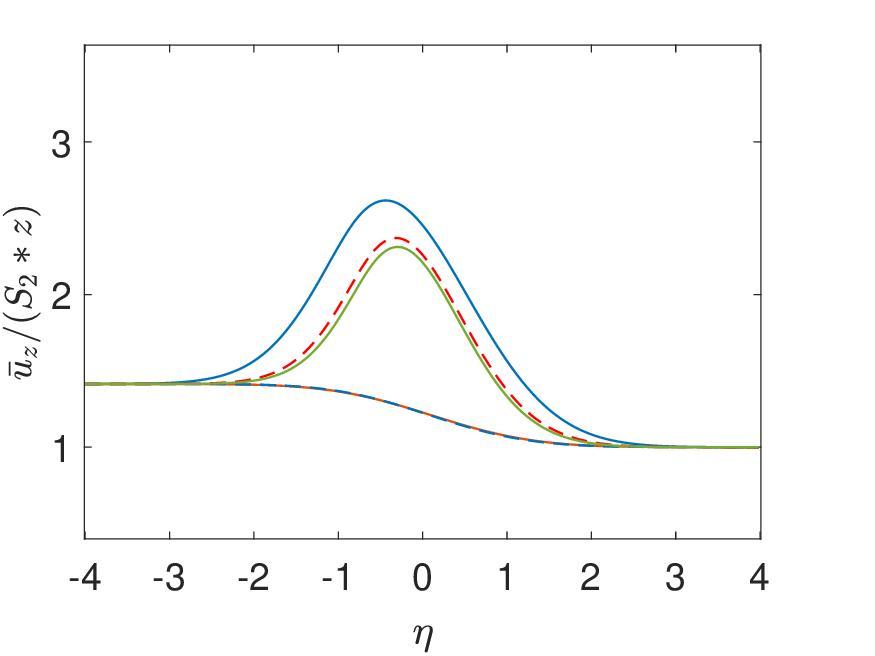}} 
  \caption{Effects of vorticity and Damk\"{o}hler number on  steady-state flow properties for diffusion flame. $S_1 =0.5, \; \sigma = 0.5,\; Pr =1.$   Solid blue, $K= 0.150, \;  \omega = 1.2$.  \;    Solid  red, $K =0.150, \;\omega = 1.0,$ \; no flame.
  \; Dashed blue, $K =0.155,\;  \omega = 0.500$  \; no flame. \;  Dashed red, $K=0.155,\; \omega = 0.707 $.  \;   Solid green, $K =0.160,\; \omega =0$.}
  \label{KOMsteady}
\end{figure}

Figure \ref{KOMsteady} shows how vorticity affects flammability especially near the limits. For the reaction-rate coefficient $K =0.160$, a strong flames exists with or without vorticity, as shown in Subfigures \ref{KOMsteady}a,b. As a decrease to $K=0.155$ occurs a transition occurs for $\omega$ between the values of 0.500 and 0.707, (i.e., $\omega^2$ between 0.250 and 0.500) whereby the flame is extinguished   below a threshold value of vorticity. The threshold becomes higher than $\omega =1.0$ when the coefficient value is reduced to $K=0.150$.
When a flame exists, Subfigures \ref{KOMsteady}c,d show that the mass flux variation is greater through the flame and the local normal strain rate $ \partial v/ \partial \chi$ becomes very large. Note that, in the figure, the derivative of $v$ with respect to $\eta$ is somewhat larger than the derivative with respect to $\chi$ because of density reduction near the flame. Still however, the stretching of the flow is substantial. 
Subfigures \ref{KOMsteady}e,f show clearly a three-dimensional flow exists with substantial influence from vorticity and burning rate. In particular, the coupling of variable density caused by the exothermic reaction with the vorticity is key becaiuse of a centrifugal effect. In fact, in the flame zone, very high vorticity is seen to cause a reversal of $u_{\xi}$ from outflow to inflow with increased outflow velocity in the $z$ direction.

\begin{figure}[thbp]
  \centering
 \subfigure[oscillatory enthalpy amplitude, real part, $\hat{H}_R$]{
  \includegraphics[height = 4.6cm]{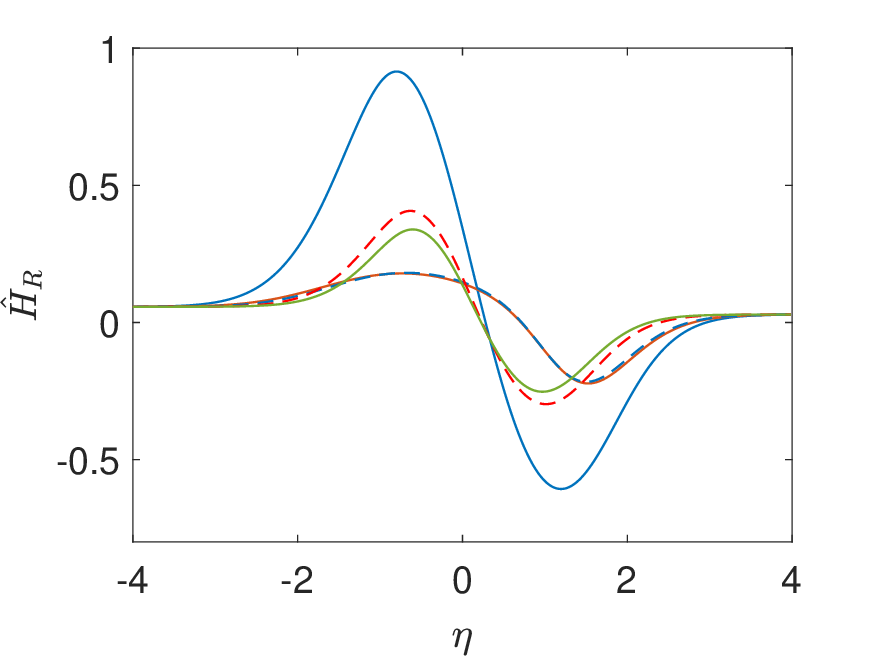}}
  \subfigure[oscillatory enthalpy amplitude, imaginary part, $\hat{H}_I$]{
  \includegraphics[height = 4.6cm]{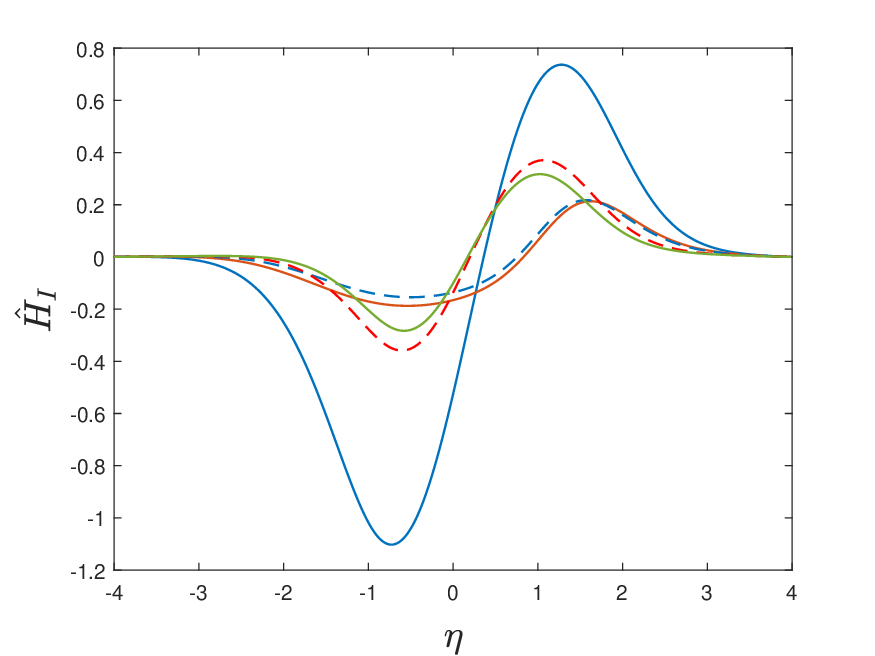}}
     \\
  \subfigure[oscillatory fuel mass fraction amplitude, real part, $\hat{Y}_{F,R}$]{
  \includegraphics[height = 4.6cm]{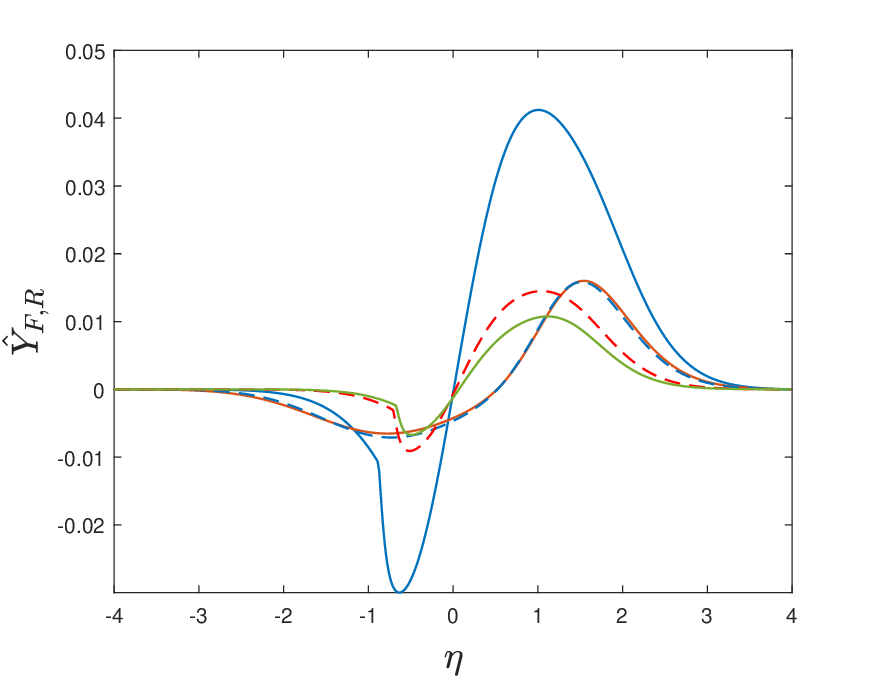}}
   \subfigure[oscillatory fuel mass fraction amplitude, imaginary part, $\hat{Y}_{F,I}$]{
  \includegraphics[height = 4.6cm]{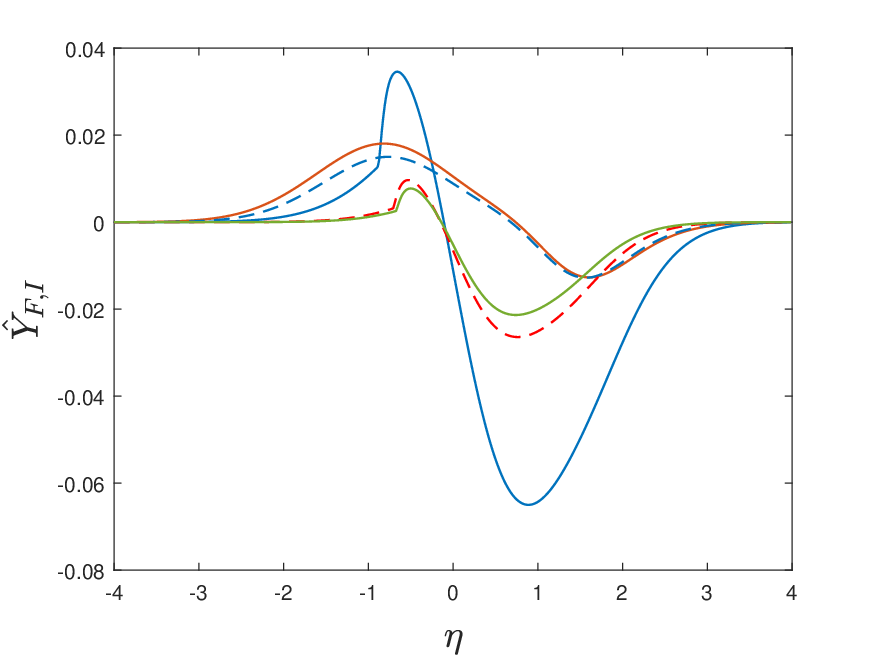}} 
  \caption{Effects of vorticity and Damk\"{o}hler number on  oscillatory enthalpy and fuel mass fraction  for diffusion flame. $S_1 =0.5, \; \sigma = 0.5,\; Pr =1.$   Solid blue, $K= 0.150, \;  \omega = 1.2$.  \;    Solid  red, $K =0.150, \;\omega = 1.0,$ \; no flame.
  \; Dashed blue, $K =0.155,\;  \omega = 0.500$  \; no flame. \;  Dashed red, $K=0.155,\; \omega = 0.707 $.  \;   Solid green, $K =0.160,\; \omega =0$.}
  \label{KOMunsteady1}
  \end{figure}
Figure \ref{KOMunsteady1} shows that, for both real and imaginary parts, the enthalpy amplitudes within the flamelet are much larger than the ambient amplitudes that are forced there by the isentropic pressure oscillation.  The oscillations in the fuel mass fraction are more gentle, presumably because the inflowing composition does not fluctuate. The amplitudes of both scalar properties increase with both vorticity $\omega$ and Damk\"{o}hler number  $Da = K Da_{ref}$.

\begin{figure}[thbp]
  \centering
 \subfigure[oscillatory $\xi$ strain-rate amplitude, real part, $\hat{F}_{1,R}$]{
  \includegraphics[height = 4.2cm]{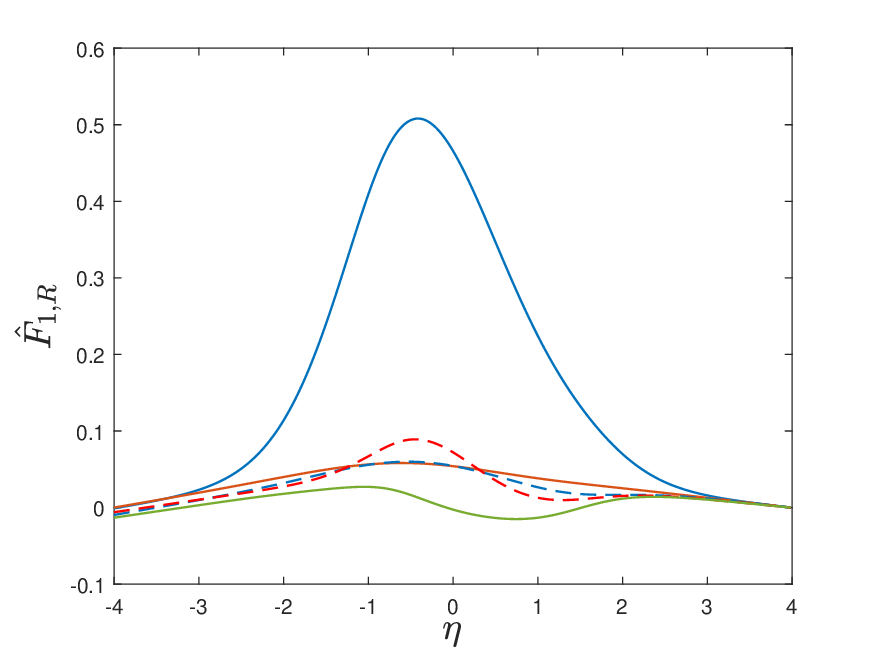}}
  \subfigure[oscillatory $\xi$ strain-rate amplitude, imaginary part, $\hat{F}_{1,I}$]{
  \includegraphics[height = 4.2cm]{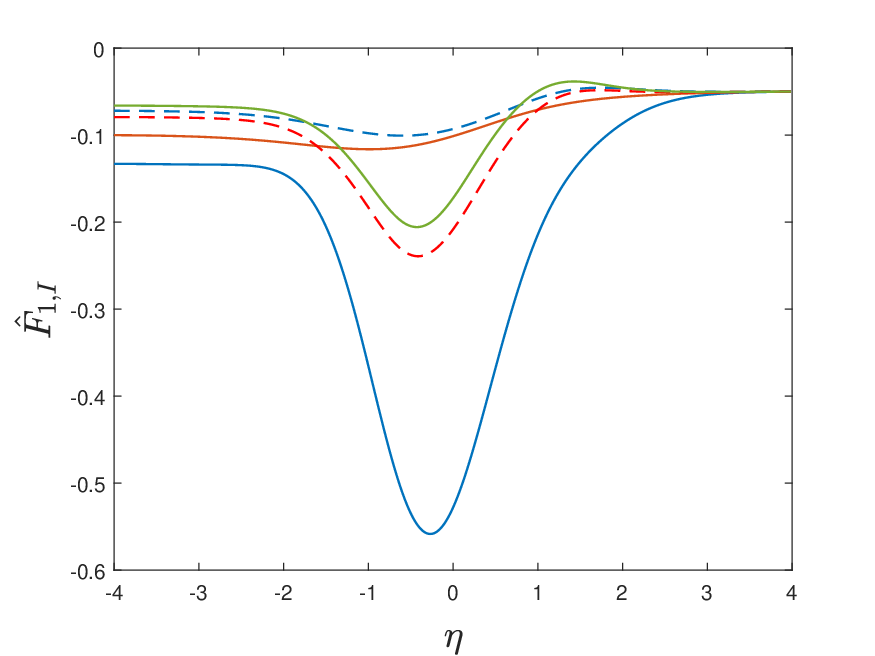}}     \\
  \subfigure[oscillatory $z$ strain-rate amplitude, real part, $\hat{F}_{2,R}$]{
  \includegraphics[height = 4.2cm]{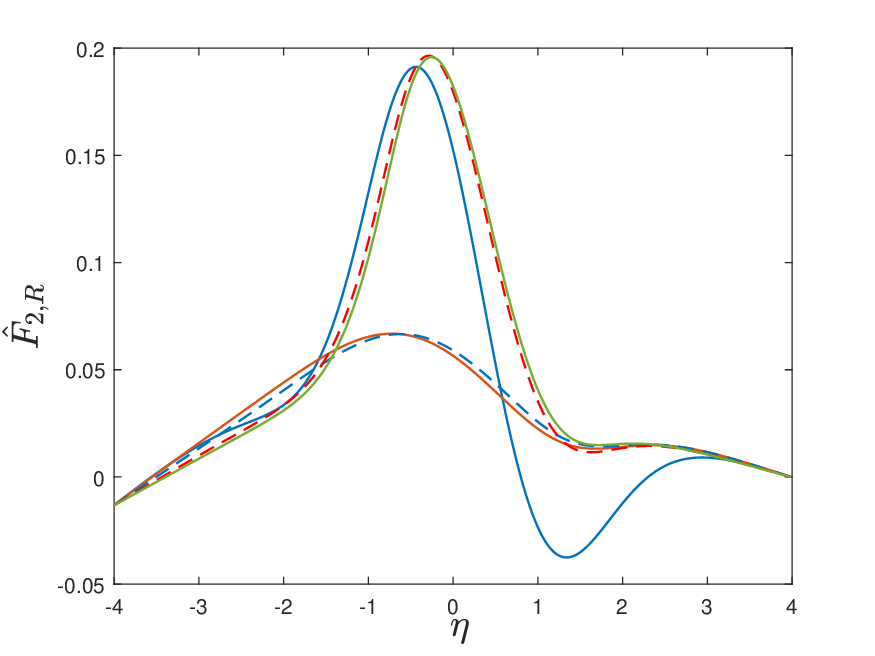}}
  \subfigure[oscillatory $z$ strain-rate amplitude, imaginary part, $\hat{F}_{2,I}$]{
  \includegraphics[height = 4.2cm]
  {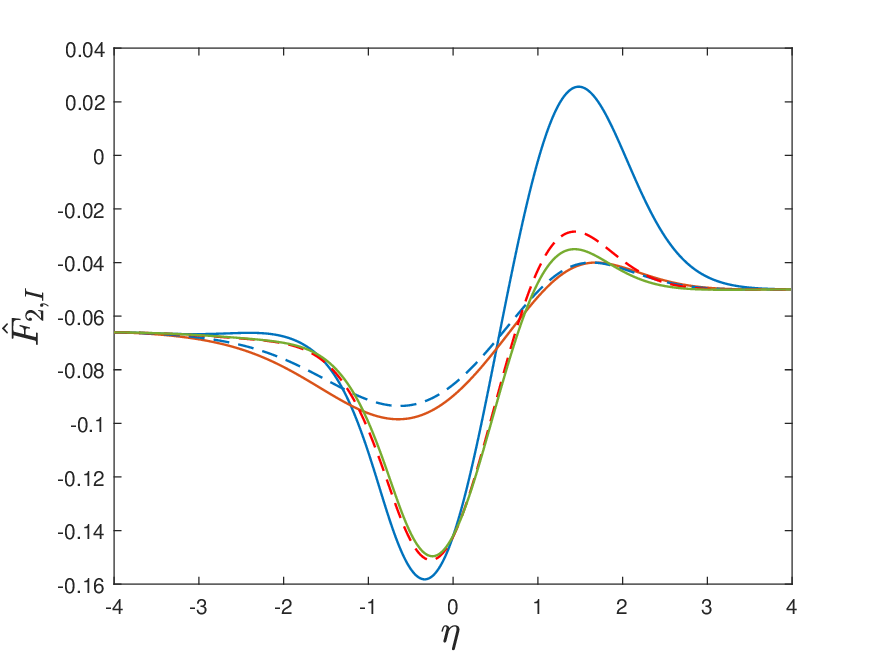}}  \\
  \subfigure[oscillatory mass-flux amplitude, real part, $\hat{f}_R$] {
  \includegraphics[height = 4.2cm]{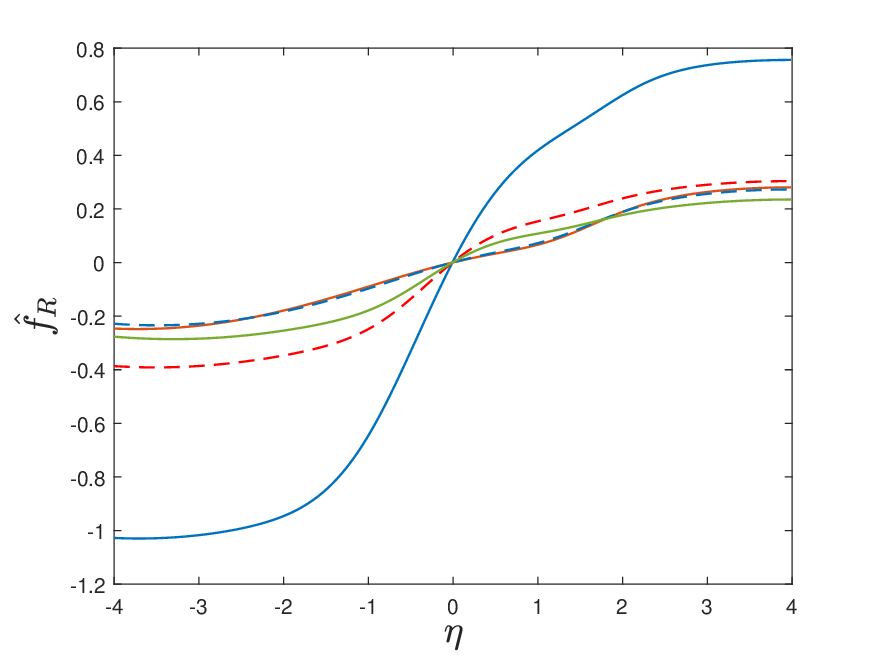}}
   \subfigure[oscillatory mass-flux amplitude, imaginary part, $\hat{f}_I$]{
  \includegraphics[height = 4.2cm]{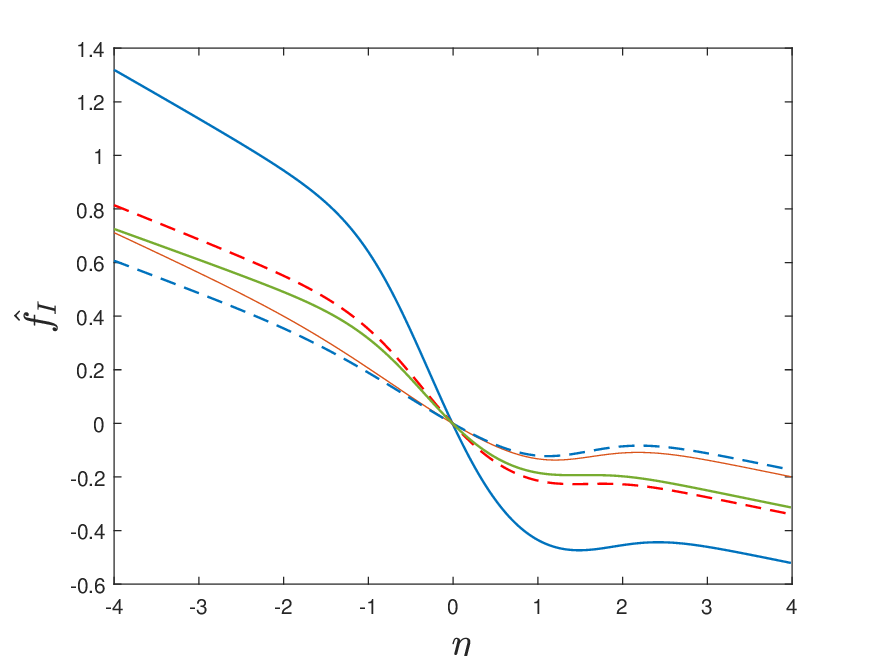}} 
  \caption{Effects of vorticity and Damk\"{o}hler number on  oscillatory strain rate and mass flux  for diffusion flame. $S_1 =0.5, \; \sigma = 0.5,\; Pr =1.$   Solid blue, $K= 0.150, \;  \omega = 1.2$.  \;    Solid  red, $K =0.150, \;\omega = 1.0,$ \; no flame.
  \; Dashed blue, $K =0.155,\;  \omega = 0.500$  \; no flame. \;  Dashed red, $K=0.155,\; \omega = 0.707 $.  \;   Solid green, $K =0.160,\; \omega =0$.}
  \label{KOMunsteady2}
  \end{figure}

Figure \ref{KOMunsteady2} demonstrates that the  normal strain rates in both the  $\xi$ and $z$ directions experience substantial amplitudes of both the real and imaginary parts. Again, the monotonically increasing relations are seen with $\omega$ and $Da$.

\begin{figure}[thbp]
  \centering
 \subfigure[oscillatory enthalpy amplitude, real part, $\hat{H}_R$]{
  \includegraphics[height = 4.6cm ]{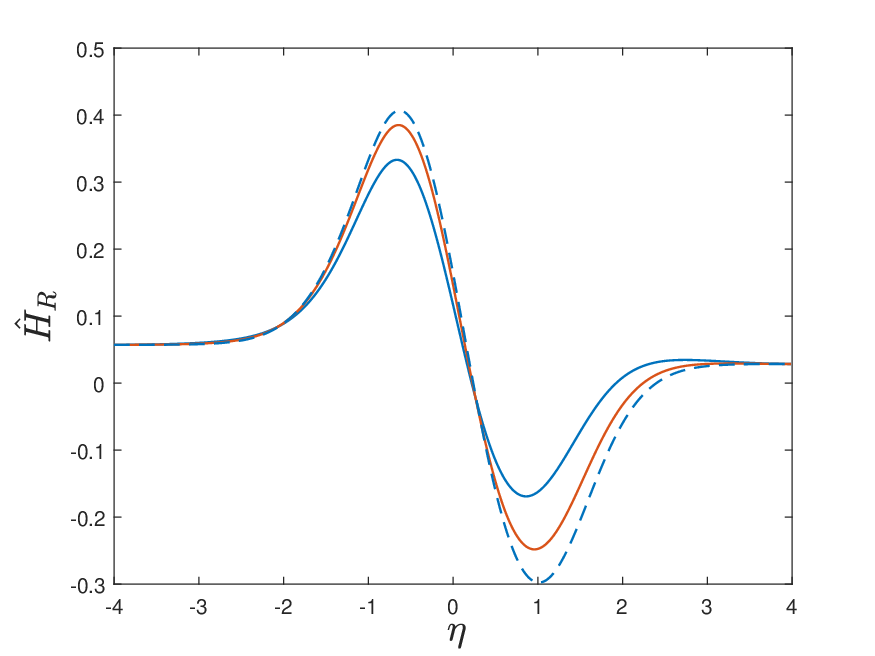}}
  \subfigure[oscillatory enthalpy amplitude,  imaginary part, $\hat{H}_I$]{
  \includegraphics[height = 4.6cm ]{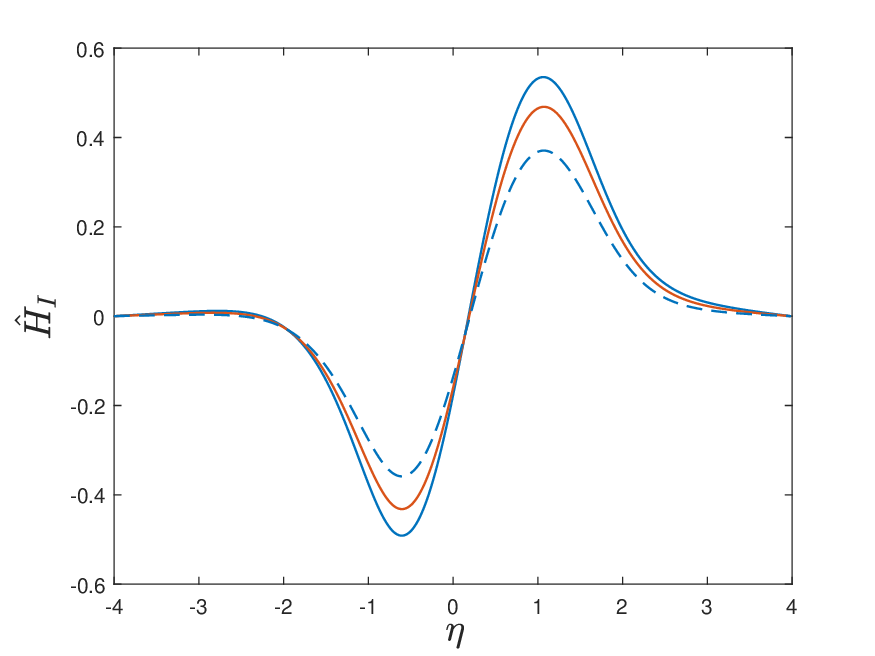}}
     \\
  \subfigure[oscillatory fuel mass fraction amplitude, real part, $\hat{Y}_{F,R}$]{
  \includegraphics[height = 4.6cm ]{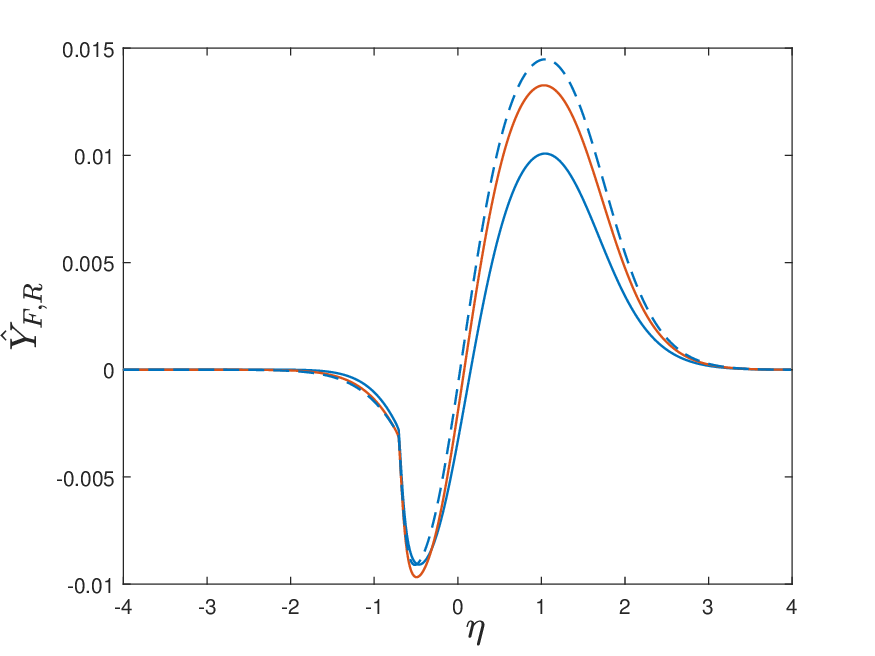}}
   \subfigure[oscillatory fuel mass fraction  amplitude, imaginary part, $\hat{Y}_{F,I}$]{
  \includegraphics[height = 4.6cm ]{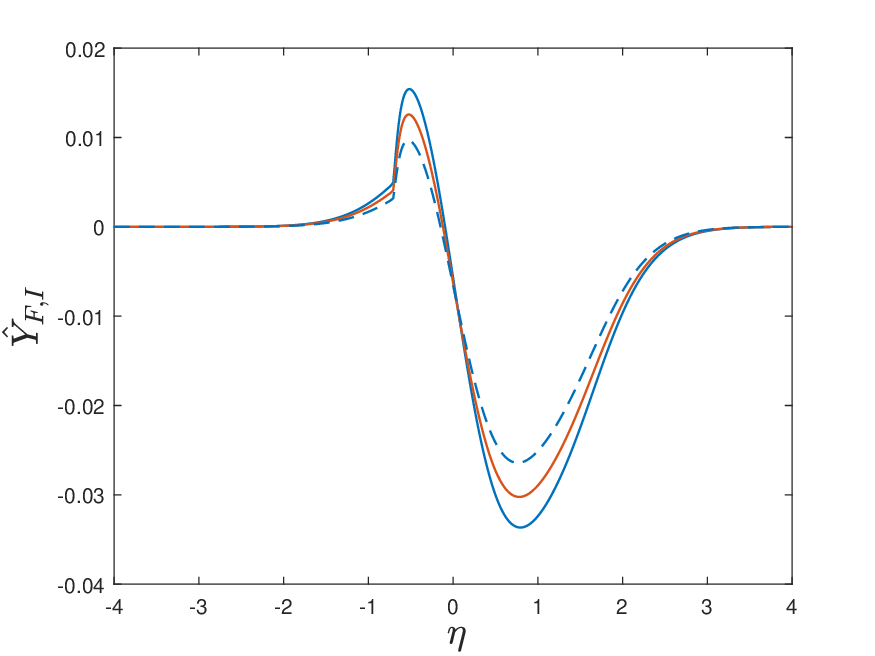}} 
  \caption{Effects of oscillation frequency on  oscillatory enthalpy and fuel mass fraction  for diffusion flame. $S_1 =0.5, \; K =0.155, \;  \omega = 0.707 ,\; Pr =1.$   Solid blue, $\sigma =1.50 $.  \;    Solid  red, $\sigma = 1.00$ .   \; Dashed blue, $\sigma = 0.50.$ }
  \label{sigunsteady1}
  \end{figure}
  
\begin{figure}[thbp]
  \centering
 \subfigure[oscillatory $\xi$ strain-rate amplitude, real part, $\hat{F}_{1,R}$]{
  \includegraphics[height = 4.6cm]{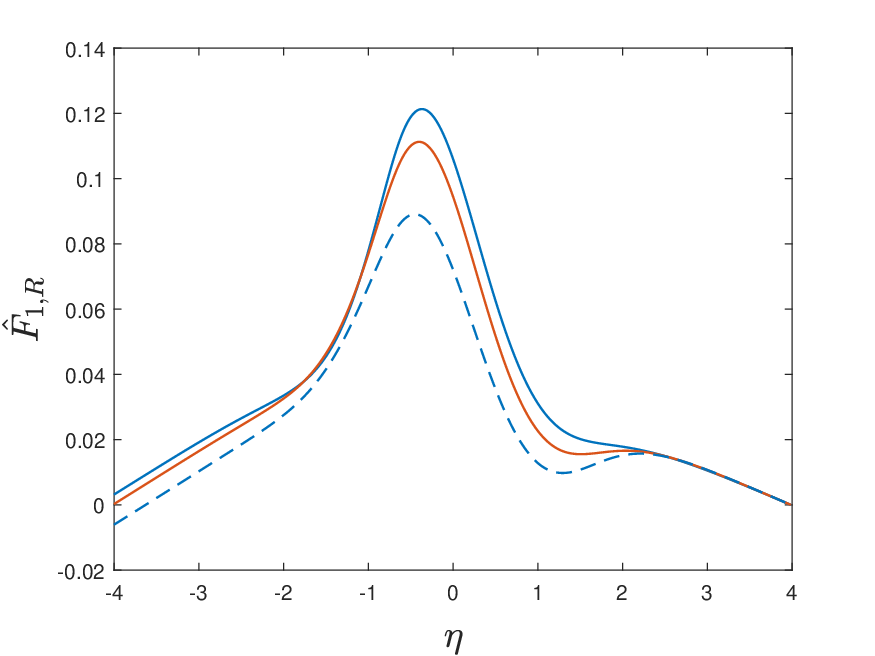}}
  \subfigure[oscillatory $\xi$ strain-rate amplitude, imaginary part, $\hat{F}_{1,I}$]{
  \includegraphics[height = 4.6cm]{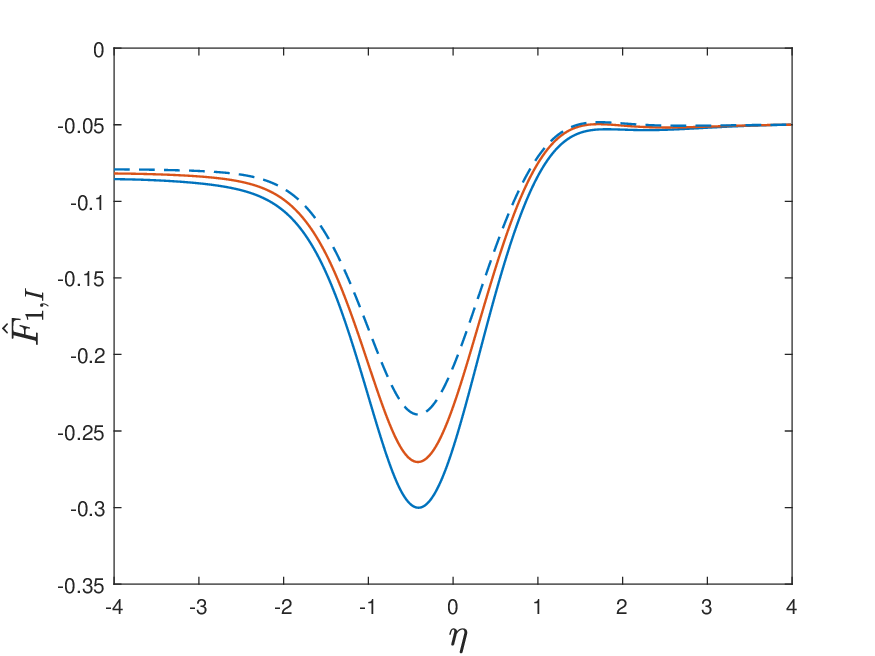}}
     \\
  \subfigure[oscillatory $z$ strain-rate amplitude, real part, $\hat{F}_{2,R}$]{
  \includegraphics[height = 4.6cm]{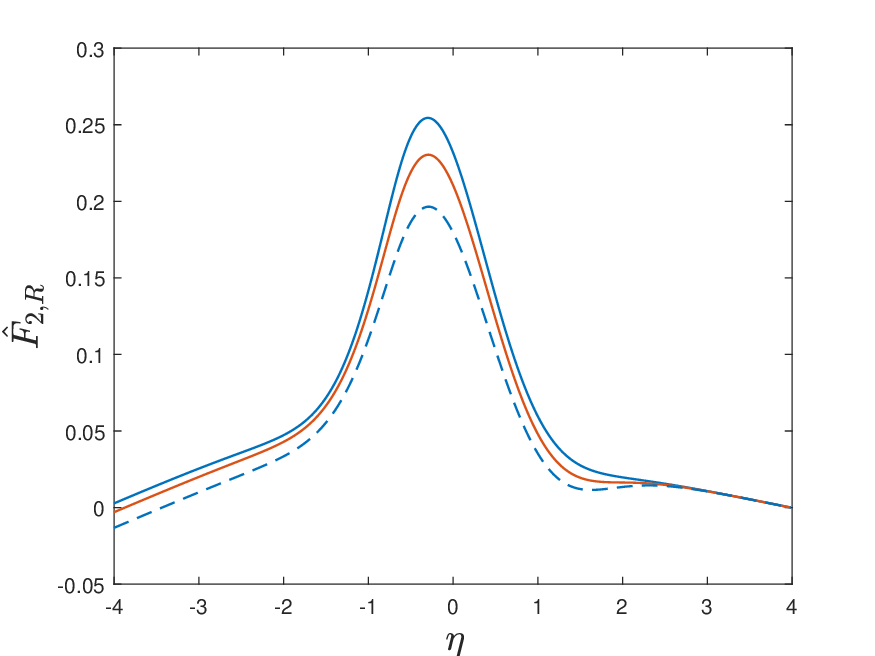}}
   \subfigure[oscillatory $z$ strain-rate amplitude, imaginary part, $\hat{F}_{2,I}$]{
  \includegraphics[height = 4.6cm]{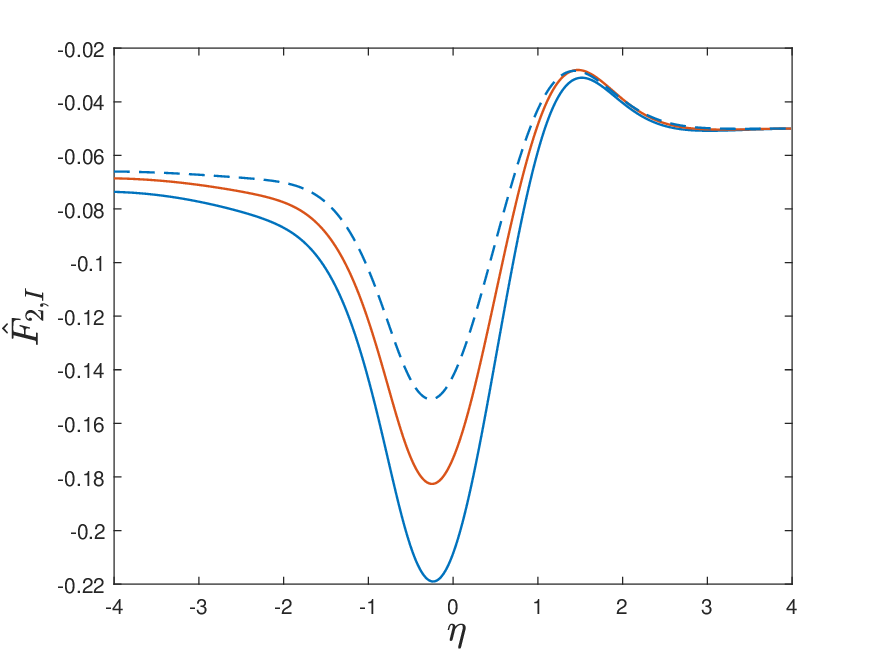}} 
  \caption{Effects of oscillation frequency on  oscillatory strain rates  for diffusion flame. $S_1 =0.5, \; K =0.155, \;  \omega = 0.707 ,\; Pr =1.$   Solid blue, $\sigma =1.50 $.  \;    Solid  red, $\sigma = 1.00$ .   \; Dashed blue, $\sigma = 0.50.$ }
  \label{sigunsteady2}
  \end{figure}

The effect of the frequency of oscillation is shown in Figures \ref{sigunsteady1} and \ref{sigunsteady2}. Frequency values  $\sigma = 0.5, 1.0,$ and $ 1.5$ are taken. The steady-state solutions for the case is shown by the dashed-red curves in Figure \ref{KOMsteady}. Amplitudes of the real and imaginary parts of strain rate and the imaginary parts of enthalpy and fuel mass fraction tend to increase with frequency. The real parts of the scalar properties decrease with increasing frequency. Interestingly, the characteristic length of the perturbation in $\chi$ or $\eta$ space does not change with frequency. Thus, a strong diffusive character is shown, as expected. versus a possible wavelike character.

\begin{figure}[thbp]
  \centering
 \subfigure[Steady-state enthalpy, $\bar{h}$]{
  \includegraphics[height = 4.6cm]{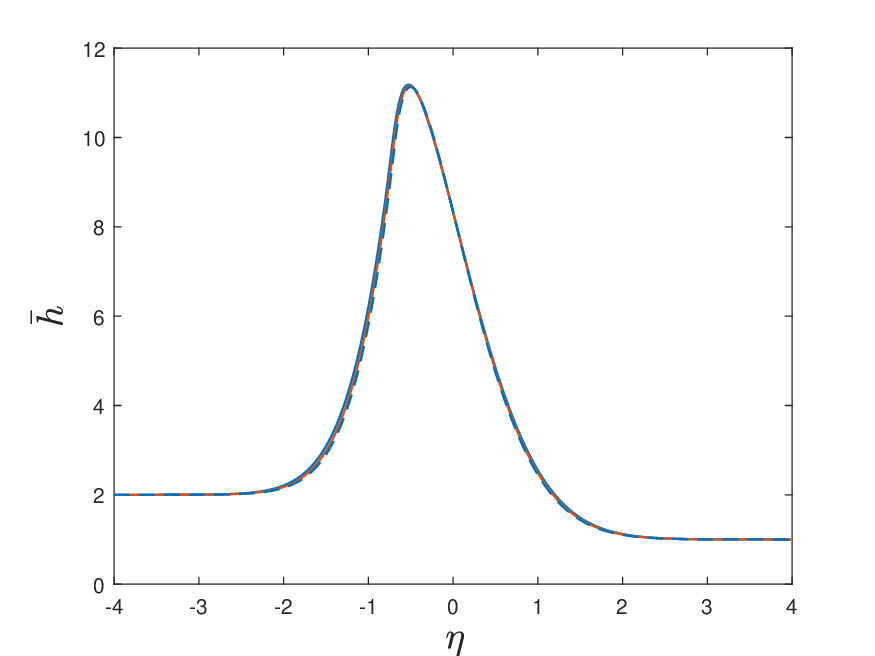}}
  \subfigure[Steady-state fuel mass fraction, $\bar{Y}_F$]{
  \includegraphics[height = 4.6cm]{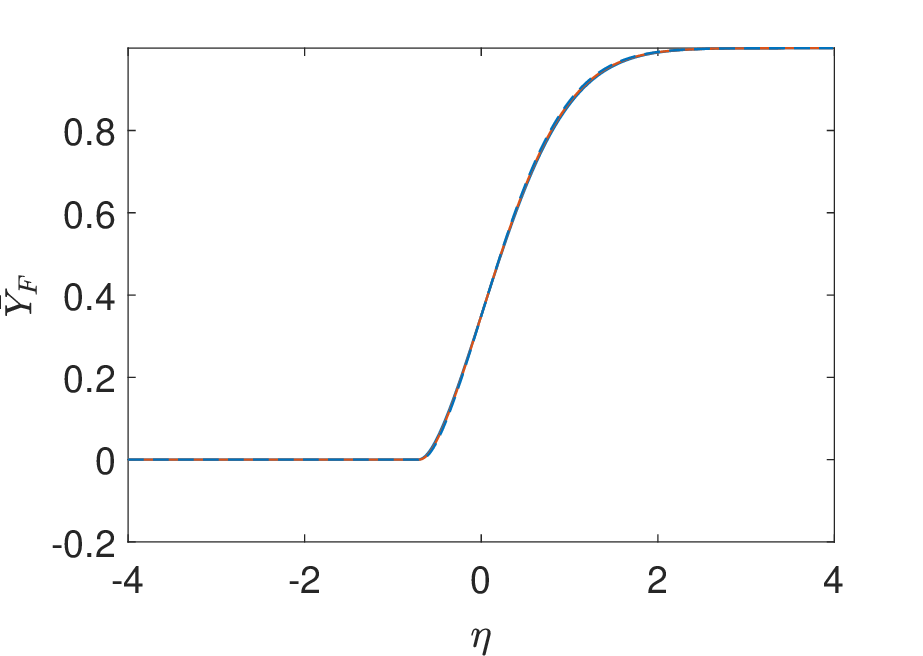}}     \\
    \subfigure[Steady-state normal strain rate $\bar{u}_{\xi}/(S_1*\xi)$]{
  \includegraphics[height = 4.6cm]{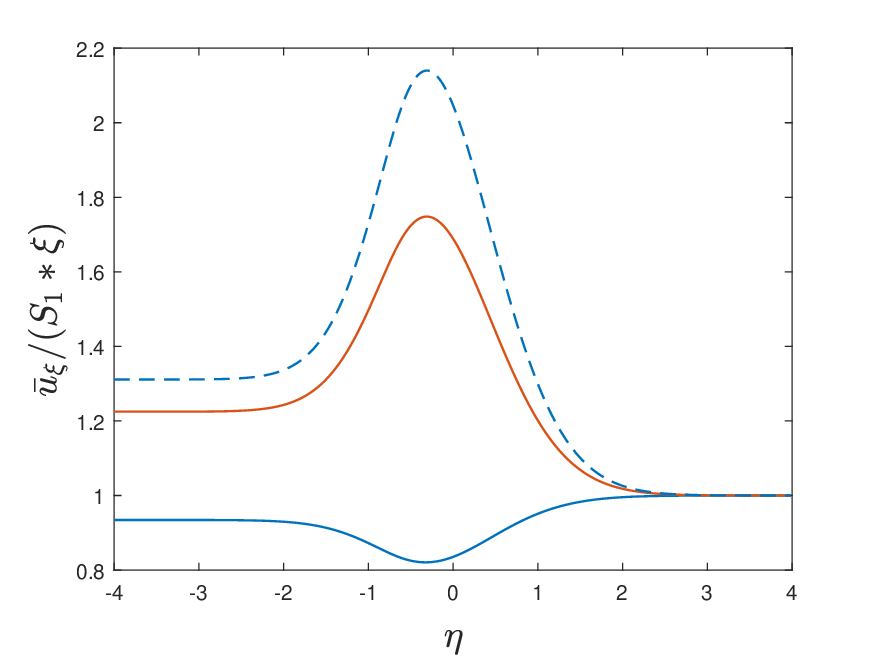}}
   \subfigure[Steady-state normal strain rate $\bar{u}_z/(S_2*z)$]{
  \includegraphics[height = 4.6cm]{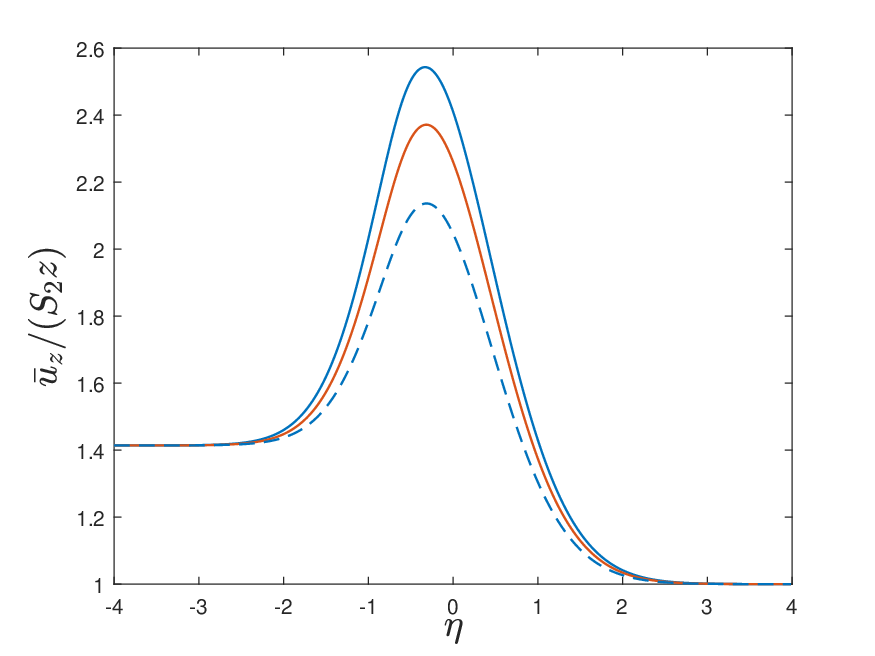}} 
  \\
  \caption{Effects of normal strain rate on  steady-state flow properties for diffusion flame. $K = 0.160, \; \omega =0.707,  \; \sigma = 0.5,\; Pr =1.$   Solid blue, $S_1 =0.333, S_2 = 0.667$.  \;    Solid  red, $S_1 = S_2 = 0.500$ . \; Dashed blue, $S_1=0.667, \; S_2 = 0.333$.}
  \label{S1S2steady}
\end{figure}
The variation of the normal strain rates $S_1$ and $S_2$, in the $\xi$ and $z$ directions respectively, have negligible effect on steady-state scalar properties as shown in Figure \ref{S1S2steady}. The steady-state  strain rate $\hat{F}_1$ increases substantially with increasing $S_1$ while $\hat{F}_2$ increases substantially with $S_2= 1-S_1$. Note that the magnitudes in the figure have been scaled by $S_1$ and $S_2$ to bring the curves closer in the graphing.

\begin{figure}[thbp]
  \centering
 \subfigure[oscillatory enthalpy amplitude, real part, $\hat{H}_R$]{
  \includegraphics[height = 4.6cm]{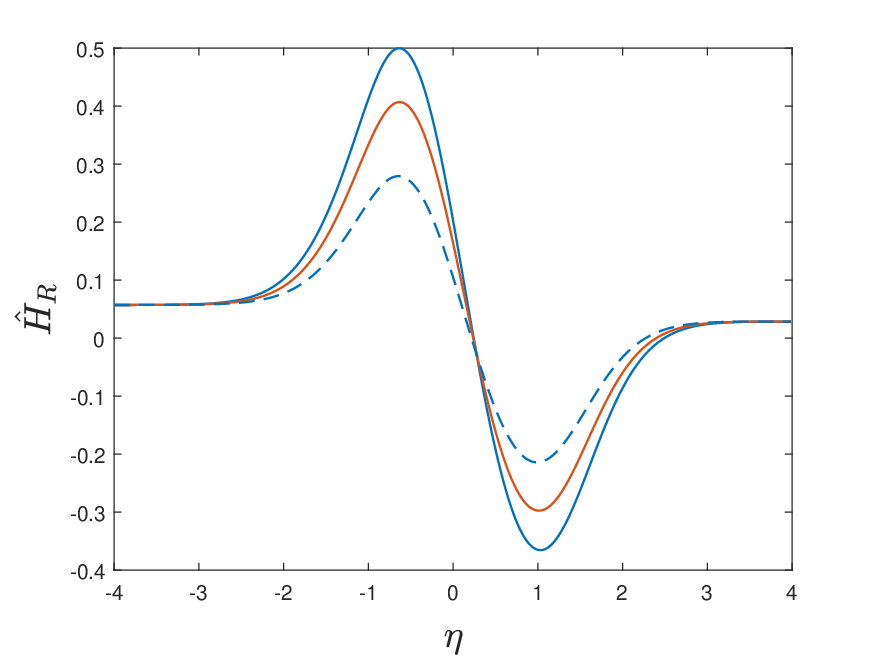}}
  \subfigure[oscillatory enthalpy amplitude, imaginary part, $\hat{H}_I$]{
  \includegraphics[height = 4.6cm]{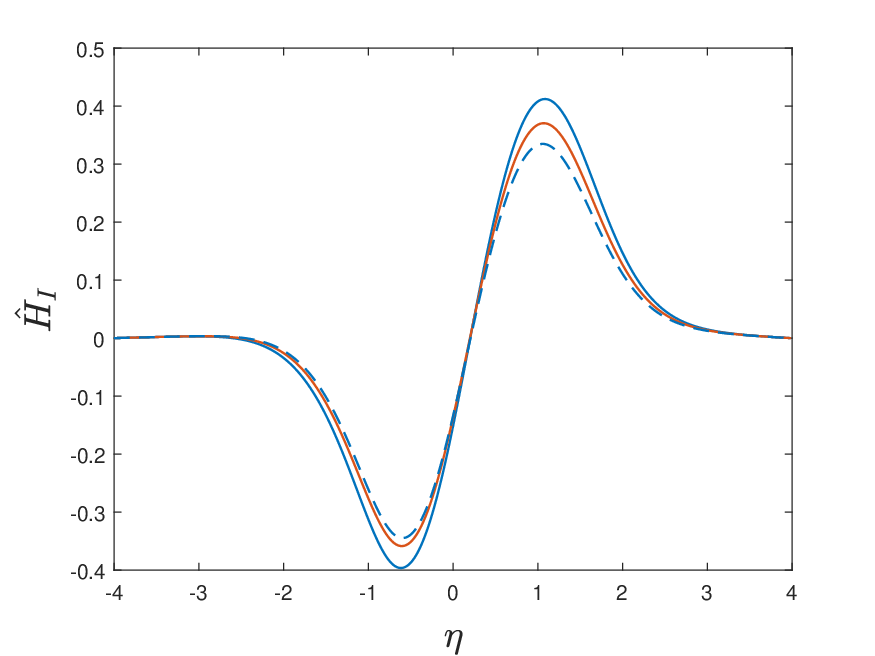}}
     \\
  \subfigure[oscillatory fuel mass fraction amplitude, real part, $\hat{Y}_{F,R}$]{
  \includegraphics[height = 4.6cm]{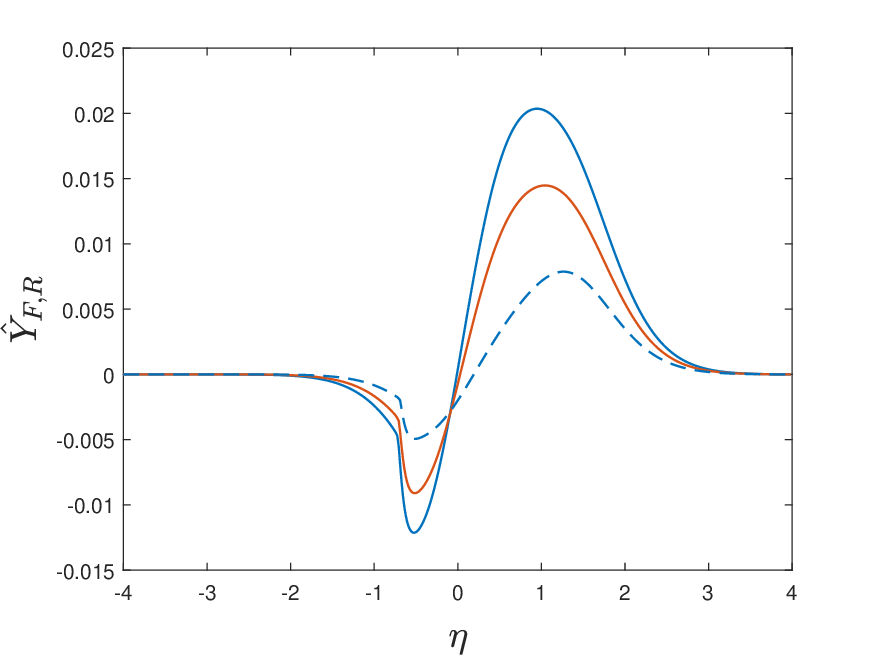}}
   \subfigure[oscillatory fuel mass fraction amplitude, imaginary part, $\hat{Y}_{F,I}$]{
  \includegraphics[height = 4.6cm]{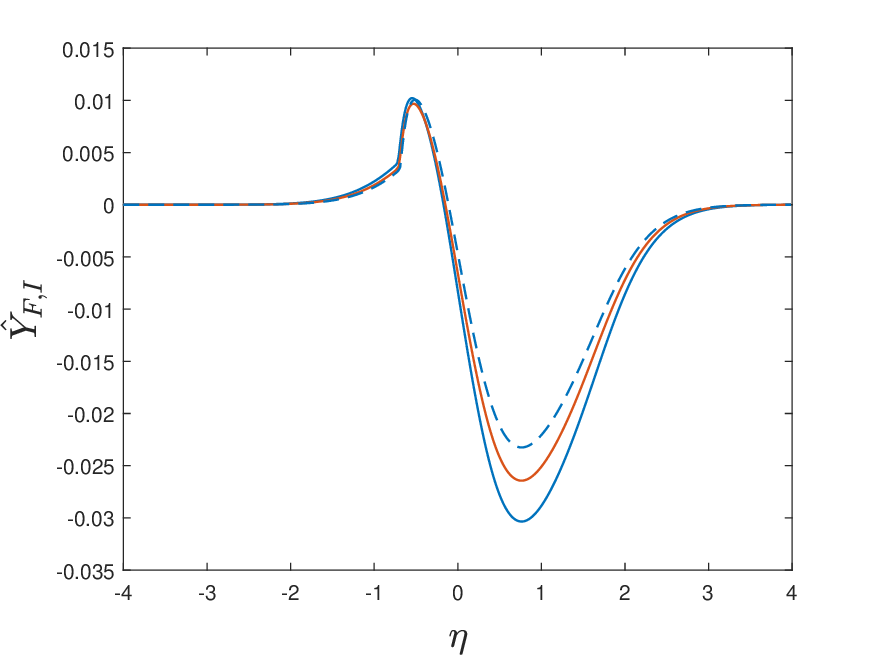}} 
  \caption{Effects of steady normal strain rates on  oscillatory enthalpy and fuel mass fraction  for diffusion flame. $K = 0.160, \; \omega =0.707,  \; \sigma = 0.5,\; Pr =1.$   Solid blue, $S_1 =0.333, S_2 = 0.667$.  \;    Solid  red, $S_1 = S_2 = 0.500$ .
  \; Dashed blue, $S_1=0.667, \; S_2 = 0.333$.}
  \label{S1S2unsteady1}
  \end{figure}

\begin{figure}[thbp]
  \centering
 \subfigure[oscillatory $\xi$ strain-rate amplitude, real part, $\hat{F}_{1,R}$]{
  \includegraphics[height = 4.6cm]{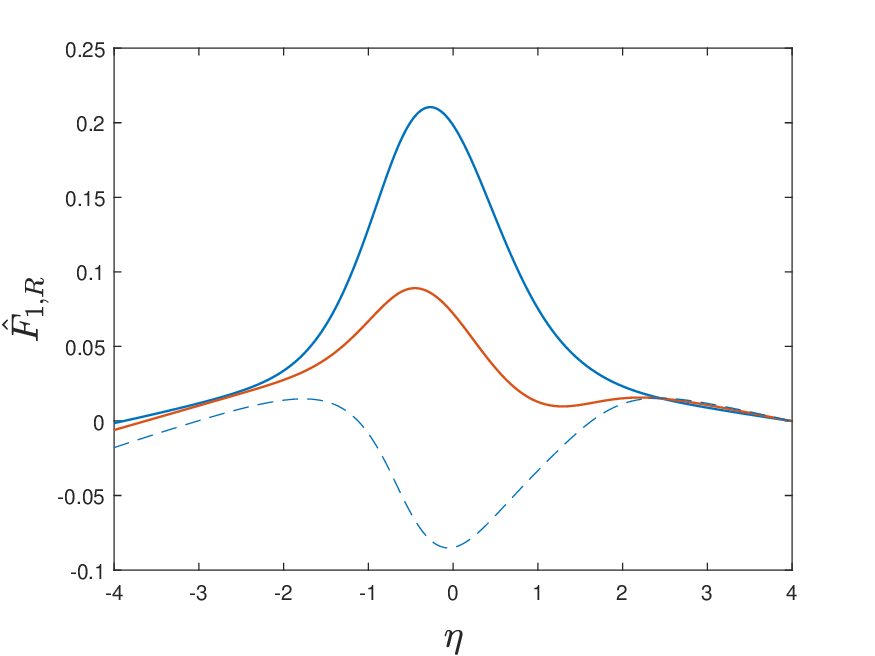}}
  \subfigure[oscillatory $\xi$ strain-rate amplitude, imaginary part, $\hat{F}_{1,I}$]{
  \includegraphics[height = 4.6cm]{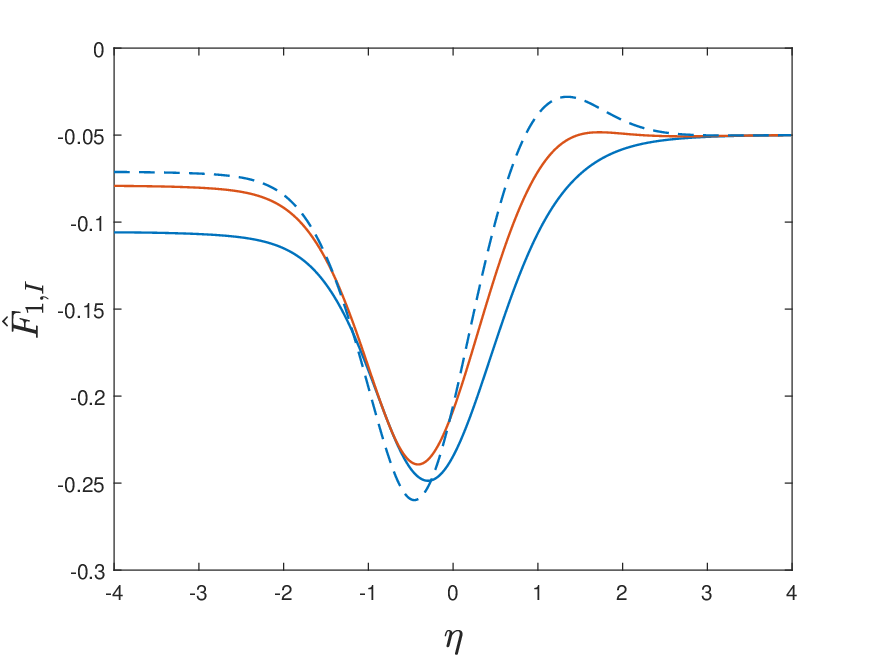}}
     \\
  \subfigure[oscillatory $z$ strain-rate amplitude, real part, $\hat{F}_{2,R}$]{
  \includegraphics[height = 4.6cm]{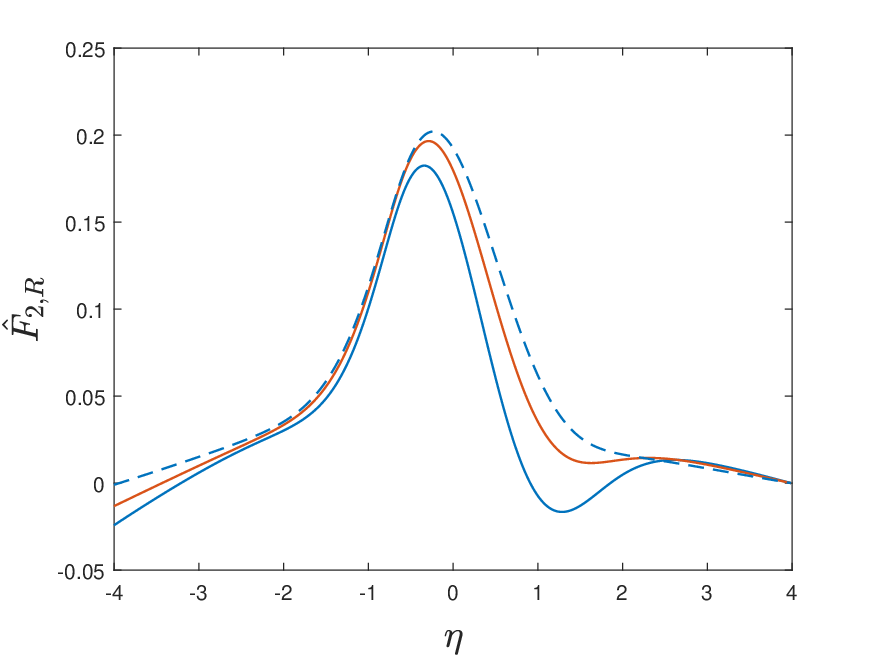}}
   \subfigure[oscillatory $z$ strain-rate amplitude, imaginary part, $\hat{F}_{2,I}$]{
  \includegraphics[height = 4.6cm]{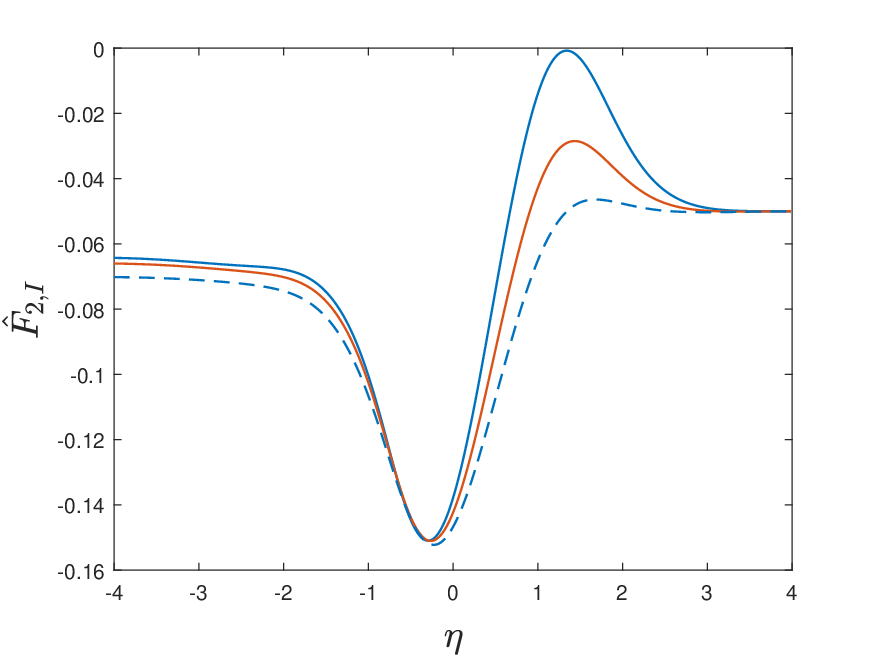}} 
  \caption{Effects of steady normal strain rates on  oscillatory strain rates  for diffusion flame. $K = 0.160, \; \omega =0.707,  \; \sigma = 0.5,\; Pr =1.$   Solid blue, $S_1 =0.333, S_2 = 0.667$.  \;    Solid  red, $S_1 = S_2 = 0.500$ .
  \; Dashed blue, $S_1=0.667, \; S_2 = 0.333$. }
  \label{S1S2unsteady2}
  \end{figure}

The impact of the applied-strain-rate distribution on the unsteady oscillatory perturbations is shown in Figures \ref{S1S2unsteady1},\ref{S1S2unsteady2}.  The characteristic length for variation of amplitude is comparable to characteristic length for steady-state variables and does not vary with $S_1$. The real and imaginary parts of the amplitudes for scalar properties and $\xi$ strain rate decreased with increasing $S_1$, while the real and imaginary parts of the $z$ strain rate increased.

\subsection{Angular Momentum} \label{omega}

 There is reason to suspect that conservation of angular momentum should be considered, especially for the time dependent problem.  We cannot always independently set the vorticity value as the ambient velocities or normal strain rates for the counterflow are chosen. 

 Let us develop a simple formulation to represent the effect of angular momentum. Consider the uniform vorticty over a volume with cross-sectional area within a disk of radius $ R = \sqrt{x^2 + y^2} = \sqrt{\chi^2 + \xi^2}$ where the disk is centered on the origin and
 the values of $\xi, \chi$ or $x, y$ are taken on the circular boundary centered at the origin.    As normal strain rate varies in time, the disk  mass  should change; that is, stretching occurs. In this model of the flow, there is no applied torque. We would expect that, without other influence, the rotational rate $\omega$, which is twice the rotational rate in radians per second, will vary so that angular momentum is conserved. Let $r = \sqrt{\xi^2 + \chi^2} \leq R$ be the distance from the origin to any point inside the circular boundary.  Since density $\rho$ does not vary with $\xi$ in our counteflow model and a symmetry exists about $\xi =0$, $\Omega$, the angular momentum $\Omega$ (per unit length in the $z$-direction) of the cylindrical domain, is given as
 \begin{eqnarray}
 \Omega &=& 2\omega \int_{-R}^{+R}\int_0^{\sqrt{R^2 -\xi^2}}\rho r^2 d\xi d\chi  \nonumber \\
   &=& 2\omega \int_{-R}^{+R}\int_0^{\sqrt{R^2 -\xi^2}}\rho (\xi^2 + \chi^2) d\xi d\chi  \nonumber \\
   &=& \frac{2}{3}\omega \int_{-R}^{+R}\rho(\chi, t)G(\chi) d\chi
\end{eqnarray}
where $G(\chi)\equiv \sqrt{R^2 -\chi^2}[R^2 +2\chi^2]$.  If $\Omega$ is constant in time, linearization yields that the perturbed vorticity
$\omega(t)$ is given by
\begin{eqnarray}
\frac{\omega'(t)}{\bar{\omega}} = - \frac{\int_{-R}^{+R}\rho'(\chi, t)G(\chi) d\chi}{\int_{-R}^{+R}\bar{\rho}(\chi)G(\chi) d\chi}
 = - e^{i\sigma t}  \frac{\int_{-R}^{+R}[\hat{\rho}(\chi)/\bar{\rho}(\chi)] \bar{\rho}(\chi)G(\chi) d\chi }{\int_{-R}^{+R}\bar{\rho}(\chi)G(\chi) d\chi}
 \label{primed}
\end{eqnarray}
So, the magnitude of $\omega'(t)$ will be bounded by $\bar{\omega} [\hat{\rho}/\bar{\rho}]_{max}$.
 Here, the modification of vorticity with time is qualitatively similar to the modification of vorticity during outflow in the steady-state stretched vortex tube described by
\cite{Burgers1948}  and \cite{Rott}. As the material particles for the steady stretched vortex flow outward in the $z$ direction, they move closer to the origin of the $x,y$ plane. Accordingly, the  vorticity value increases in order to maintain the circulation magnitude. As the normal strain rate is increased for the steady stretched vortex, the stretching effect is increased, moving material particles still closer to the $x, y$ plane origin and increasing the magnitude of the vorticity. In our unsteady model here, as the normal compressive strain rate increases (decreases) with time, the material particles are pressed closer to (farther from) the $x, y$ origin, thereby increasing (decreasing) the vorticity in time.

The formula given by Equation (\ref{primed}) identifies issues more than it resolves an issue.  There is uncertainty about the magnitude of $R$ and its maximum value where the constant $\omega$ value remains reasonable. These questions are left for future study.

$\omega'$ would explicitly appear only in differential equation or boundary conditions for the perturbed variables $F'_1, u'_{\xi}, $ and $\partial p'/\partial \xi$. However, the implicit impact on all other variables exists because it affects $f'$ and $\rho'$.
 
A simplified formula for $\omega'\equiv \hat{\omega}exp(i\sigma t)$ can be created by realizing that the integrand weighting increases with the square of the distance from the origin.  Replacement of $[\hat{\rho}(\chi)/\bar{\rho}(\chi)]$ by
$[\hat{\rho}(\infty)/\bar{\rho}(\infty)]/2 + [\hat{\rho}(-\infty)/\bar{\rho}(-\infty)]/2 = (1/\gamma)\hat{p}/\bar{p}$ leads to a possible approximation $\omega' \approx \bar{\omega} \hat{p}\; exp(i\sigma t) /(\gamma \bar{p})$.

\begin{figure}[thbp]
  \centering
 \subfigure[oscillatory enthalpy amplitude, real part, $\hat{H}_R$]{
  \includegraphics[height = 4.6cm]{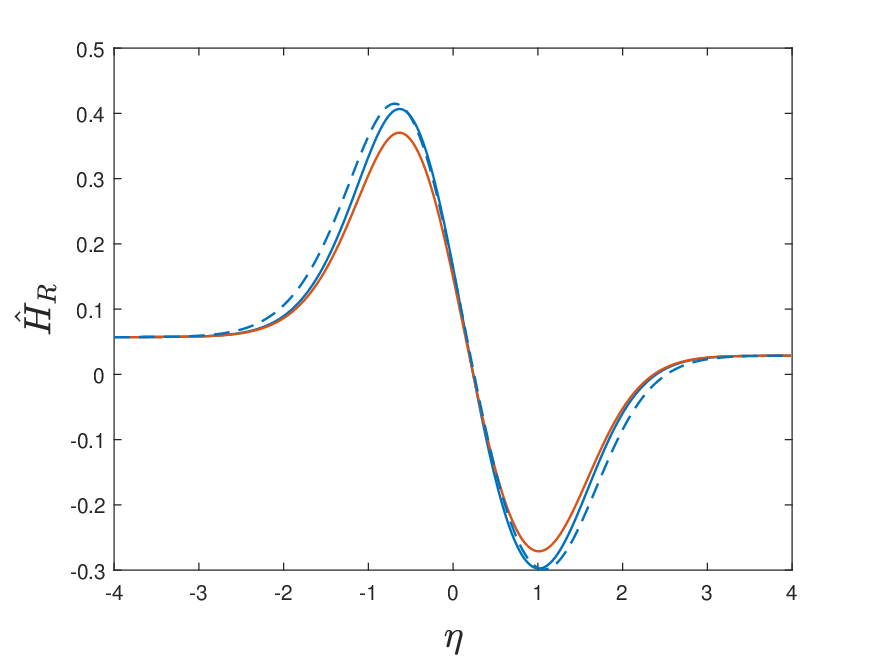}}
  \subfigure[oscillatory enthalpy amplitude,  imaginary part, $\hat{H}_I$]{
  \includegraphics[height = 4.6cm]{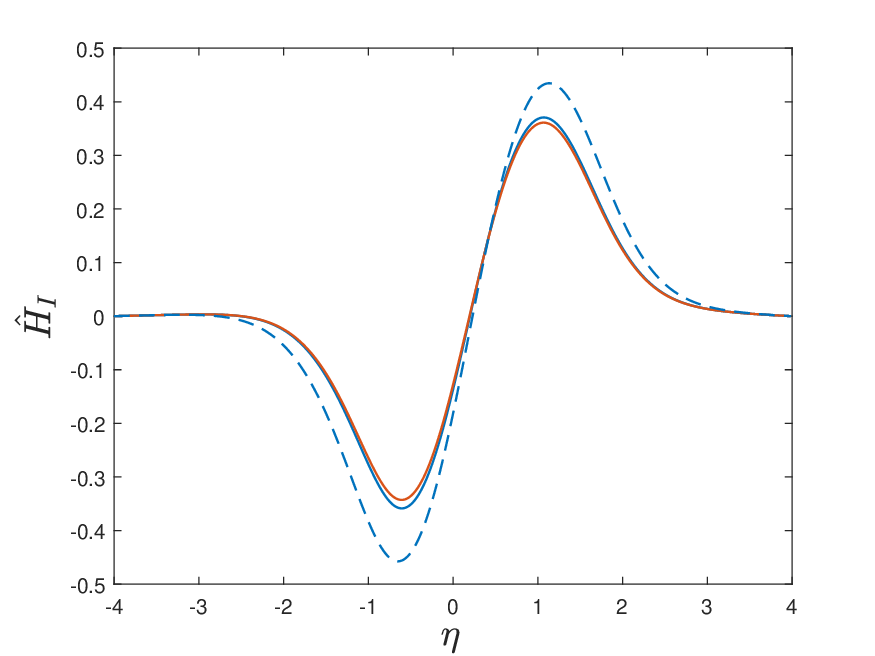}}
     \\
  \subfigure[oscillatory fuel mass fraction  amplitude, real part, $\hat{Y}_{F,R}$]{
  \includegraphics[height = 4.6cm]{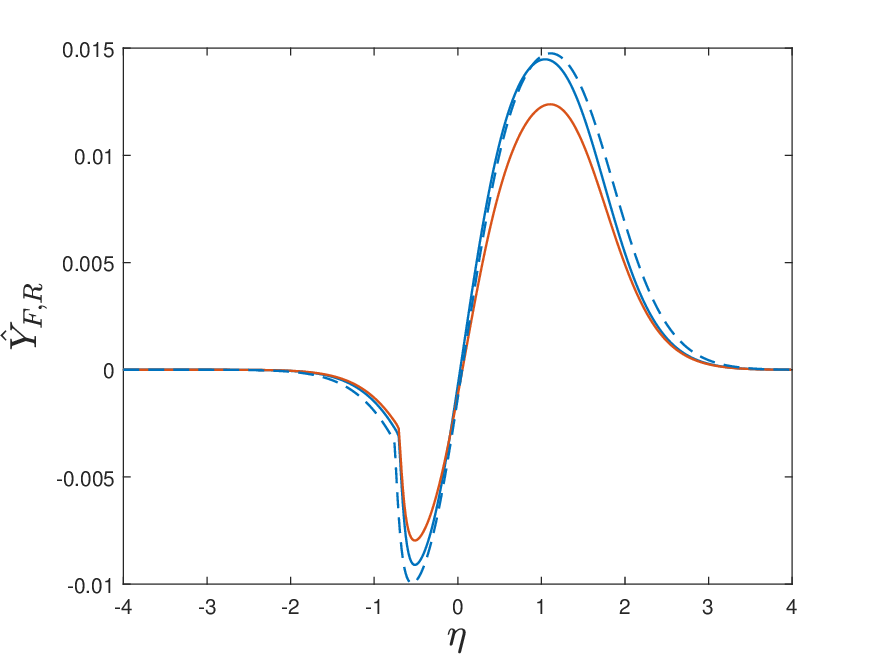}}
   \subfigure[oscillatory fuel mass fraction  amplitude, imaginary part, $\hat{Y}_{F,I}$]{
  \includegraphics[height = 4.6cm]{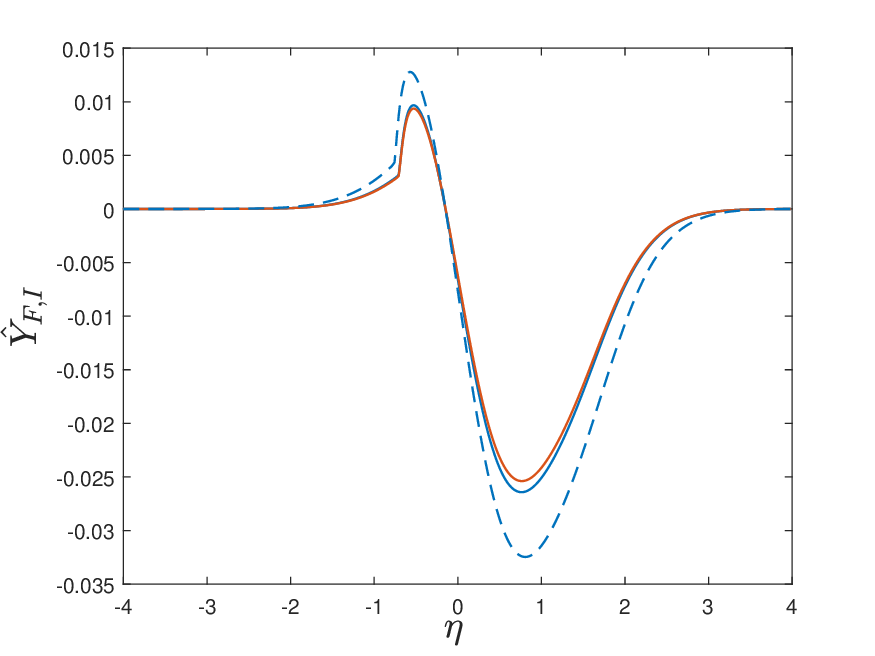}} 
  \caption{Effects of vorticity perturbation on  oscillatory enthalpy and fuel mass fraction  for diffusion flame.  $S_1 =0.5, \; K =0.155, \;   \hat{p} =0.1, \; \sigma =0.5, \; Pr =1.$   Solid blue, $\bar{\omega} =  0.707,$ \; with  no vorticity perturbation.  \;    Solid  red, $\bar{\omega} =0.707$ with perturbation.   \; Dashed blue, $\bar{\omega} = 1.0$ with perturbation.  }
  \label{AMunsteady1}
  \end{figure}

\begin{figure}[thbp]
  \centering
 \subfigure[oscillatory $\xi$ strain-rate amplitude, real part, $\hat{F}_{1,R}$]{
  \includegraphics[height = 4.6cm]{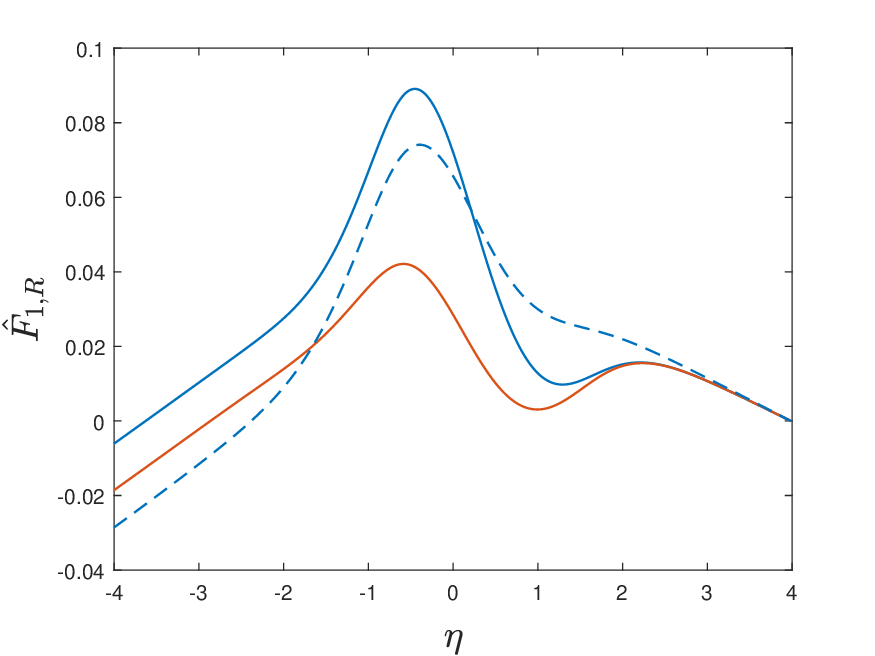}}
  \subfigure[oscillatory $\xi$ strain-rate amplitude, imaginary part, $\hat{F}_{1,I}$]{
  \includegraphics[height = 4.6cm]{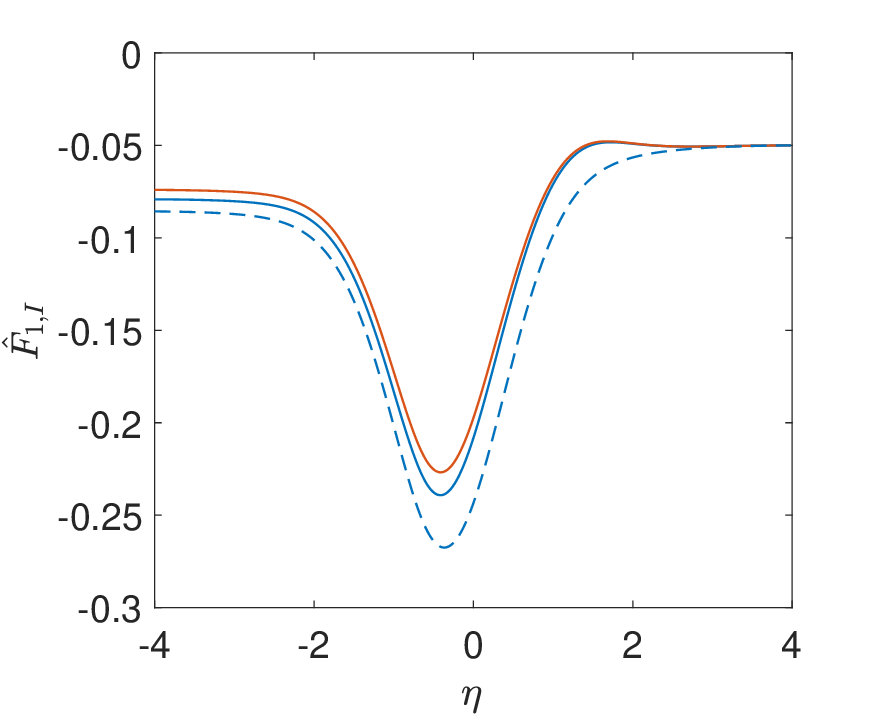}}
     \\
  \subfigure[oscillatory $z$ strain-rate amplitude, real part, $\hat{F}_{2,R}$]{
  \includegraphics[height = 4.6cm]{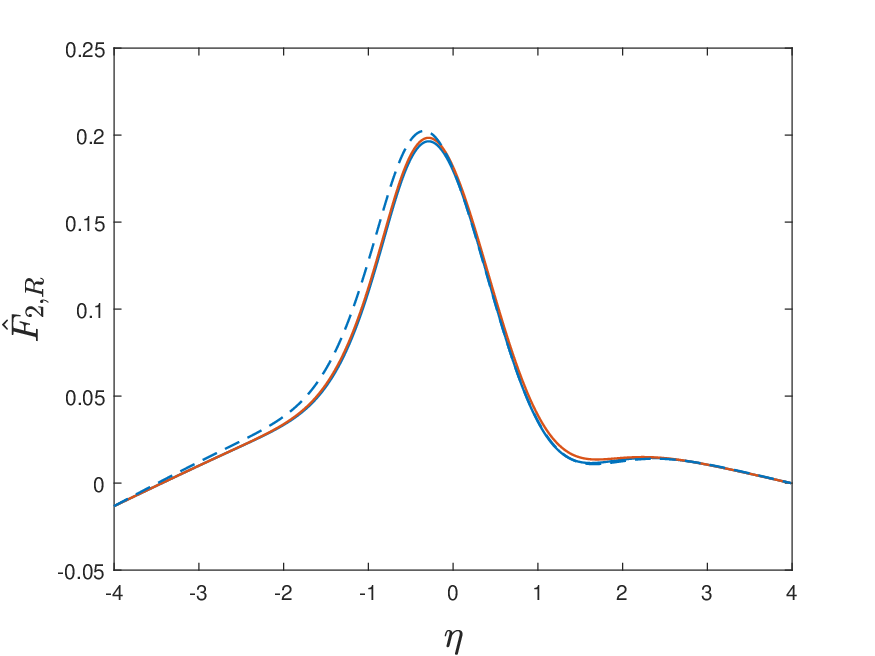}}
   \subfigure[oscillatory $z$ strain-rate amplitude, imaginary part, $\hat{F}_{2,I}$]{
  \includegraphics[height = 4.6cm]{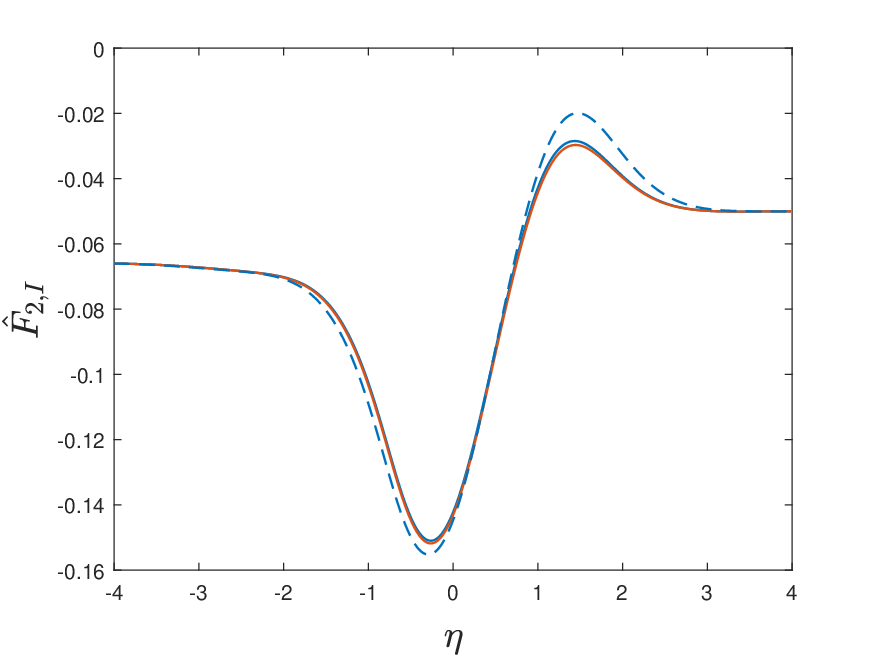}} 
  \caption{Effects of vorticity perturbation on  oscillatory strain rates  for diffusion flame. $S_1 =0.5, \; K =0.155, \;   \hat{p} =0.1, \; \sigma =0.5, \; Pr =1.$   Solid blue, $\bar{\omega} =  0.707,$ \; with  no vorticity perturbation.  \;    Solid  red, $\bar{\omega} =0.707$ with perturbation.   \; Dashed blue, $\bar{\omega} = 1.0$ with perturbation. }
  \label{AMunsteady2}
  \end{figure}

Here, we require the addition of one term to the first differential equation given for $F_1(\eta)$ in the set of Equation (\ref{perturbationODEs}). Specifically, $+[1/\bar{\rho} -1]\bar{\omega}\hat{\omega}/2$ is added to the right side. We also modify in consistent fashion the boundary condition for $F_{1, -\infty}$. The calculations given in Figures \ref{AMunsteady1} and \ref{AMunsteady2} use $K =0.155, \; \sigma = 0.5,\; S_1 = S_2 = 0.5, \; \hat{p} = 0.1$, with $\bar{\omega} = 0.707$ for the two solid curves  and $\bar{\omega} = 1.00$  for the dashed curves. Thus, the steady-state results with $\bar{\omega} = 0.707$ are identical to those given by the dashed red curves in Figure \ref{KOMsteady}.   The effect of the vorticity perturbation $\omega'$ on the oscillatory perturbations of the scalar properties is seen to be modest in Figure \ref{AMunsteady1}; it is most pronounced for the real part of the fuel mass fraction perturbation. The average vorticity $\bar{\omega}$ does have impact as seen in earlier results.  The effect on the oscillatory perturbations of strain rate are shown in \ref{AMunsteady2} to have strong effects on the real part of the $\xi$ normal strain rate amplitude with modest effects on other terms.  Generally, it appears that the neglect of any effect of $\omega'$ in the calculations presented in \ref{oscillation} has generally (but not always) small 
quantitative consequences with no sign of qualitative misrepresentation.

\section{Concluding Remarks}\label{conclusions}

A new unsteady flamelet model is developed for use in  analysis of turbulent combustion. This flamelet model presents certain key advances: (i) the impacts of shear strain and vorticity (and associated centrifugal effects) on the  flames are determined; (ii) the applied sub-grid strain rates and vorticity have potential to be related to the resolved-scale strain rates and vorticity without the use of a contrived progress variable; (iii) the flamelet model is three-dimensional without need for assuming axisymmetry or planar geometry, allowing the physically correct counterflow under the vorticity constraint; (iv) variable density is addressed in the flamelet model. non-premixed flames; and (v) nonpremixed flames, premixed flames, or multi-branched flame structures are allowed to appear naturally without prescription.  Each of these five features introduces consequential, vital physics that is missed by current two-dimensional, irrotational, constant-density flamelet models that assume a priori a nonpremixed- or premixed-flame structure and make no direct connection to shear strain or vorticity on the larger turbulence scales.

Information from direct numerical simulations concerning the relative alignments of the vorticity vector, scalar gradients, and principal strain axes provides a basis for a set of assumptions.  In the steady-state limit, this unsteady analysis reduces to the quasi-steady, rotational  flamelet model of \cite{Sirignano_2022a, Sirignano_2022b}.

The order of the three-dimensional, unsteady problem has been reduced under certain conditions with fluctuating strain rates and vorticity. Specifically, a one-dimensional, time-dependent similar solution is given for the flamelet model.  Nevertheless, all three velocity components are resolved. The rotation due to vorticity creates a centrifugal force that generally decreases the mass-flux through the flamelet counterflow. Thereby, an increase in residence time and a decrease in burning rate occur. Rotation can thereby allow flames, which  would otherwise extinguish, to survive.  

 The unsteady formulation is novel whether the flamelet is rotational or irrotational. Proper formulation of the inflow boundary conditions for the similar solutions are identified, especially  when the time-dependent densities of the two inflowing streams are not identical.  The unsteady relations between  the strain rates for the two, incoming, opposing streams of the flamelet counterflow have been established and shown to be critical.  The proper boundary conditions for both the unsteady rotating flamelet and the unsteady irrotational flamelet have been presented.  A comparison with incompressible, inviscid rotational counterflow has been made to reveal certain physics about the interfacial behavior.  The conditions have been  discussed  under which the interface of the two inflows can be assumed to be fixed or approximately fixed. The opposing effects of counterflow and vorticity on the curvature of the pressure function have been analyzed and discussed.   A role of angular momentum conservation has been identified to determine vorticity fluctuation caused by strain-rate fluctuation.  Analysis and computational results for a flamelet with imposed oscillations have been presented and explained.

Several future studies developments can be built on the analysis. The existing analysis allows or studies of premixed and partially premixed flames and for the effects of non-unitary Prandtl number. Modest analytical adjustments can allow for relaxation of the unitary Lewis number assumption.  The use of detailed chemical kinetics, differential diffusion, improved descriptions of thermophysical properties, and study of the unstable burning branch can be made following \cite{Hellwig2023}. The analysis and computations can be extended to address non-oscillatory and nonlinear unsteady perturbations to the steady-state flow.  Further examinations of counterflow-interface motion and temporal change of vorticity due to angular momentum conservation would also be interesting and useful.

\section*{Acknowledgements}

 The effort was supported by AFOSR through Award  FA9550--22-1-0191
 managed by Dr. Mitat Birkan and now by  Dr. Justin Koo and by ONR through Award N00014-21-1-2467
 managed by Dr. Steven Martens.

\section*{Declaration of Interest}

The author has no known competing financial interests or personal relationships that could appear to have influenced the work reported in this paper.

\bibliographystyle{jfm}
\bibliography{Revised3DCompress}

\end{document}